\documentclass[prd,twocolumn,amsmath,amssymb,floatfix,superscriptaddress]{revtex4-1}

\usepackage{graphicx}
\usepackage{amssymb}
\usepackage{amsmath}
\usepackage{color}
\usepackage{ wasysym }
\usepackage{hyperref}
 \usepackage{tabu}

\def\barray{\begin{array}}
\def\earray{\end{array}}
\def\be{\begin{equation}}
\def\ee{\end{equation}}
\def\ben{\begin{equation} \nonumber}
\def\een{\end{equation}}
\def\ban{\begin{eqnarray*}}
\def\ean{\end{eqnarray*}}
\def\ba{\begin{eqnarray}}
\def\ea{\end{eqnarray}}

\def\({\left(}
\def\){\right)}

\def\aap{A\&A}
\def\apj{ApJ}

\def\mnras{MNRAS}
\def\araa{ARA\&A}
\def\aj{AJ}

\def\physrep{Phys. Rep.}

\def\apjs{ApJS}

\def\prd{Phys. Rev. D}

\def\jcap{J. of Cosmology and Astroparticle Physics}

\def\ssr{Space Science Reviews}

\graphicspath{{./fig/}}

\begin{document}

\title{Spherical collapse model in agegraphic dark energy cosmologies }
\author{Mehdi Rezaei}
\email{m.rezaei92@basu.ac.ir}
\affiliation{Department of Physics, Bu-Ali Sina University, Hamedan
65178, 016016, Iran}

\author{Mohammad Malekjani}
\email{malekjani@basu.ac.ir}
\affiliation{Department of Physics, Bu-Ali Sina University, Hamedan
65178, 016016, Iran}

\begin{abstract}
Under the commonly used spherical collapse model, we study how dark energy affects the growth of large scale structures of the Universe in the context of agegraphic dark energy models. The dynamics of
the spherical collapse of dark matter halos in nonlinear regimes is determined by the properties of the dark energy model. We show that the main parameters of the spherical collapse model are directly affected by the evolution of dark energy in the agegraphic dark energy models. We compute the spherical collapse quantities for different values of agegraphic model parameter $\alpha$ in two different  scenarios:
first, when dark energy does not exhibit fluctuations on cluster scales,  and second, when dark energy inside the overdense region collapses similar to dark matter. Using the Sheth-Tormen and Reed mass functions, we investigate the abundance of dark matter halos in the framework of agegraphic dark energy cosmologies. The model parameter $\alpha$ is a crucial parameter in order to count the abundance of dark matter halos. Specifically, the present
analysis suggests that the agegraphic dark energy model with bigger (smaller) value of $\alpha$ predicts less (more)
virialized halos with respect to that of $\Lambda$CDM cosmology.  We also show that in agegraphic dark energy models, the number  of halos strongly depends on clustered or uniformed distributions of dark energy. 
\end{abstract}
\maketitle

\section{Introduction}
In the past two decades, various observational data gathered by different independent cosmological experiments including those of type Ia 
supernova \citep{Riess1998,Perlmutter1999,Kowalski2008},
cosmic microwave background (CMB)
\citep{Komatsu2009,Jarosik:2010iu,Komatsu2011,Ade:2015yua},
large scale structure and baryonic acoustic oscillation
\citep{Tegmark:2003ud,Cole:2005sx,Eisenstein:2005su,Percival2010,Blake:2011rj,Reid:2012sw},
high redshift galaxies \citep{Alcaniz:2003qy}, high redshift galaxy
clusters \citep{Wang1998,Allen:2004cd} and weak gravitational
lensing \citep{Benjamin:2007ys,Amendola:2007rr,Fu:2007qq} indicate that our Universe is undergoing a period of cosmic acceleration.There is a gap in our understanding of the cosmological dynamics. In fact we do not understand the cause and nature of this accelerated expansion. To interpret this phenomenon, cosmologists follow two approaches. Some believe that this acceleration reflects on the physics of gravity at cosmological scales. They are trying to modify general relativity (GR) to justify the accelerated expansion of the Universe. In this way they propose modified gravity models that have been wildly studied in the literature, e.g.,  $f(R)$ gravity \citep{Buchdahl:1983zz},the Randall-Sundrum
model \citep{Randall:1999vf}, the Dvali-Gabadadze-Porrati (DGP) model  \citep{Dvali:2000hr} , the generalized braneworld model \citep{Dvali:2003rk}, and the modified DGP model \citep{Koyama:2006ef}. On the other hand, one can adopt GR and try to justify the accelerated expansion by introducing a new form of fluid with sufficiently negative pressure named dark energy(DE). Based on the latest observational experiments, this unknown fluid occupies about $70 \%$ of the total energy budget of the Universe \citep{Ade:2015yua}. Einstein cosmological constant $\Lambda$  with constant EoS parameter $w_{\rm \Lambda}=-1$ is the first and simplest candidate for DE. However, the standard $\Lambda$ cosmology suffers from severe theoretical problems the so-called fine-tuning and cosmic coincidence problems \citep{Weinberg1989,Sahni:1999gb,Carroll2001,Padmanabhan2003,Copeland:2006wr}. The $\Lambda$ problems persuade cosmologists to seek a DE model with a time-varying equation of state parameter. Recently, several attempts on this way led to the appearance of new dynamical DE models with a time varying EoS parameter  proposed extensively in literature. Quintessence\citep{Caldwell:1997ii,Erickson:2001bq}, ghost \citep{Veneziano1979,Witten1979,Kawarabayashi1980,Rosenzweig1980}, holographic \citep{Horava2000,Thomas2002}, k-essence\citep{Armendariz2001}, tacyon\citep{Padmanabhan2002}, chaplygin gas\citep{Kamenshchik:2001cp}, generalized chaplygin gas\citep{Bento:2002ps}, dilaton \citep{Gasperini2002,Arkani2004,Piazza2004}, phantom\citep{Caldwell2002}, quintom\citep{Elizalde:2004mq} are examples of such dynamical DE models. In this work we focus on the agegraphic dark energy model (see section \ref{sect:NADE}) as  the most interesting model in the family of dynamical DE models.

More deeply speaking, DE not only causes the accelerated expansion of the Universe, but also affects the scenario of structure formation in late times. It is believed that 
the large scale structures in the Universe  are developed from the gravitational collapse of primordial small density perturbations\citep{Gunn1972,Press1974,White1978,Peebles1993,Peacock1999,Peebles2003,Ciardi2005,Bromm2011}. Initial seeds of these density perturbations are produced during the phase of  inflationary expansion \citep{Guth1981,Linde1990}.
An analytical and simple approach for studying the evolution of matter fluctuations is the spherical collapse model (SCM) , first introduced by \cite{Gunn1972}.
In this scenario, due to self-gravity, spherical overdense regions expand slower compared with Hubble flow . Therefore the overdense sphere becomes denser and denser (compare to background). At a certain redshift the so-called turnaround redshift, $z_{\rm ta}$,  the overdense sphere completely decouples from the background fluid and starts to collapse. The collapsing sphere finally reaches the steady state at a virial radius in certain redshift $z_{\rm vir}$. SCM in standard and DE cosmologies has been wildly investigated in several works\citep{Fillmore1984,Bertschinger1985,Hoffman1985,Ryden1987,Subramanian2000,Ascasibar2004,Williams2004,Mehrabi:2016exz}. It has been extended for various cosmological models\citep{Mota2004,Maor2005,Basilakos:2006us,Abramo2007,Schaefer:2007nf,Abramo2009a,Li2009,Pace2010,Pace2012,Naderi2015,Nazari-Pooya:2016bra,Malekjani:2015pza}. In this work we investigate the SCM in agegraphic dark energy cosmologies and predict the abundance of virialized halos in this model. The paper is organized as follows: In sec. \ref{sect:NADE} we introduce the agegraphic dark energy and describe the evolution of Hubble flow in this model. In sec. \ref{growth}, the basic equations for evolution of density perturbations in linear and nonlinear regimes are presented. In sec. \ref{sec:mass} we compute the predicted mass function  and cluster number in our model  in both clustered and homogeneous DE approaches.
Finally in sec.\ref{conclude} we conclude and summarize our results.

\section{Hubble flow in agegraphic DE}\label{sect:NADE}

In this section we review DE models that are constructed based on the holographic principle
in the quantum gravity scenario \citep{tHooft1993,Susskind1995}.
According
to the holographic principle, the number of degrees of freedom of
a finite-size system should be finite and bounded by the area of
its boundary \citep{Cohen1999}.
If we have a system with size $L$, its  total energy should not exceed the mass of a black hole
with the same size, i.e.,$ L^3 \rho_{\rm \Lambda} \leq L {m_{\rm p}}^2$, where $\rho_{\rm \Lambda}$ is the quantum
zero-point energy density caused by UV cutoff  and $m_{\rm p}$ is the
Planck mass ($ m_{\rm p} = 1/8\pi G$). In cosmological contexts, when
the whole of the Universe is taken into account, the vacuum energy related
to the holographic principle can be viewed as a DE, the so-called
holographic dark energy (HDE) with energy density given by

\begin{equation}\label{HDE}
\rho_{\rm d}=3 \alpha^2 m_{\rm p}^2 L^{-2}\;,
\end{equation}

where $\alpha$ is a positive numerical constant and the coefficient 3 is for convenience. It should be noted that the HDE model is defined by assuming
an IR cutoff $L$ in Eq.(\ref{HDE}). One of the choices for IR cutoff
is the Hubble length, $L= H^{-1}$. In this case, DE density will be close to the observational data, but 
the current accelerated expansion of the Universe cannot be recovered
\citep{Horava2000,Cataldo2001,Thomas2002,Hsu2004}. Another choice for the IR cutoff is the particle horizon,
which however, does not lead to the current accelerated expansion \citep{Horava2000,Cataldo2001,Thomas2002,Hsu2004}.The final choice for $L$ is to use the event horizon
\citep{ Li2004}. By choosing event horizon, not only can the HDE
justify the accelerated expansion of the Universe, but also it is consistent
with observations \citep{Pavon2005,Zimdahl2007,Sheykhi:2011cn}. Based on the holographic principle as well as using the Karolyhazy relation the authors of Refs.\citep{Karolyhazy:1966zz,Karolyhazy1982,Cai2007} suggest the new model agegraphic dark energy (ADE) in which the length scale $L$ is replaced by cosmic time $T$. Karolyhazy and Lukacs \cite{Karolyhazy:1966zz,Karolyhazy1982} made an interesting
observation concerning the distance measurement for Minkowski
spacetime through a light-clock Gedanken experiment. They found
that the distance $t$ in Minkowski spacetime cannot be known to a
better accuracy than $\delta t = \beta{t_{\rm p}}^{2/3}t^{1/3}$, where $ \beta$ is a dimensionless
constant of order $O(1)$ . Based on
the Karolyhazy relation, Maziashvili \cite{Maziashvili:2007vu} argued that the energy
density of metric fluctuations in the Minkowski spacetime is
given by  $\rho_{\rm d} \sim \frac{1}{t^2 {t_{\rm p}}^2} \sim \frac{m_{\rm p}}{t^2}$,
where $m_{\rm p}$ and $t_{\rm p}$ are the reduced Planck mass and the Planck time, respectively \citep[see also][]{Sasakura:1999xp, Ng:1993jb, Ng:1995km,
Krauss:2004fb,Christiansen:2005yg,Arzano:2006wp}. Using this form for $\rho_{\rm d}$, Cai \cite{Cai2007} proposed 
the ADE model in which the time scale $t$ is chosen to be equal with $T$, the age of the Universe. The ADE energy density
is given by \citep{Cai2007}

\begin{equation}\label{ADE}
\rho_{\rm d}=\dfrac{3 \alpha^2 m_{\rm p}^2}{T^2} \;.
\end{equation}

\begin{figure} 
	\centering
	\includegraphics[width=0.5\textwidth]{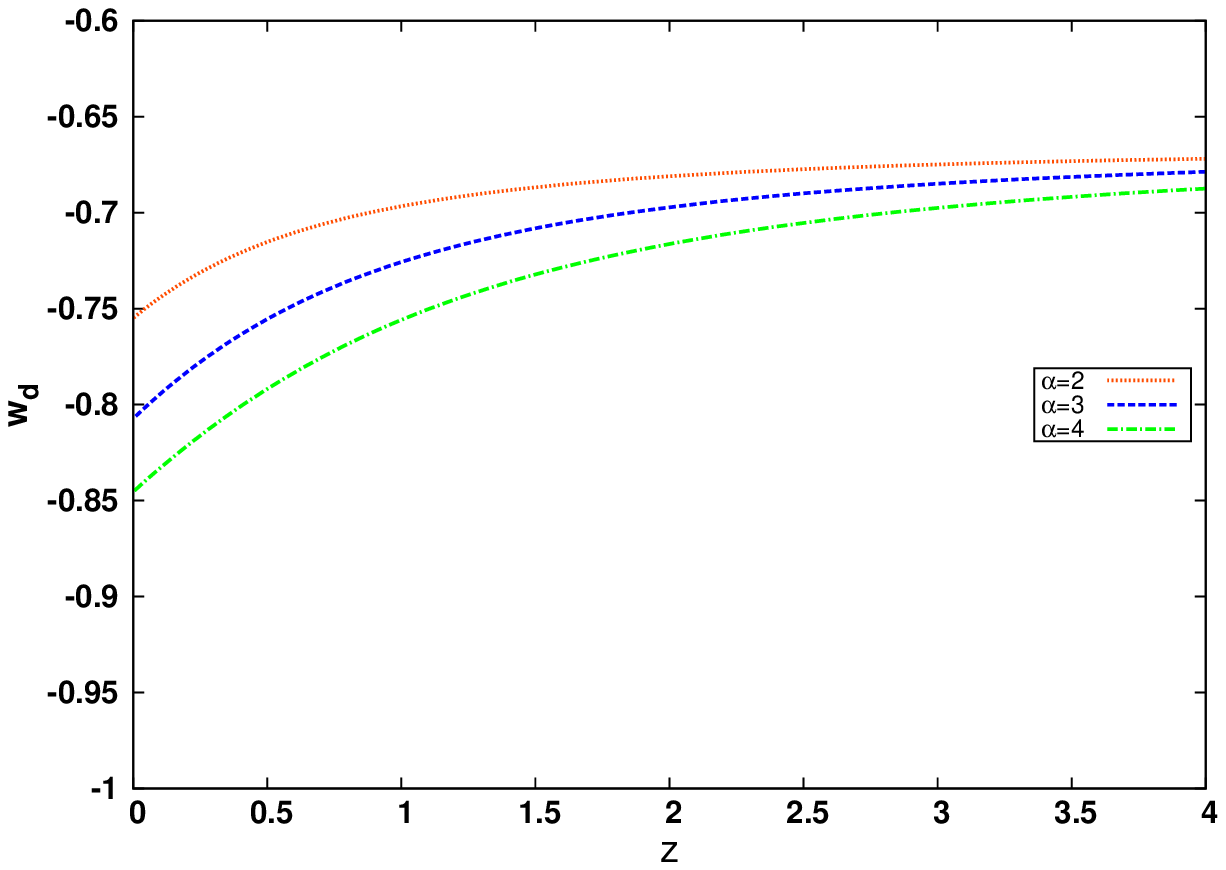}
	\includegraphics[width=0.5\textwidth]{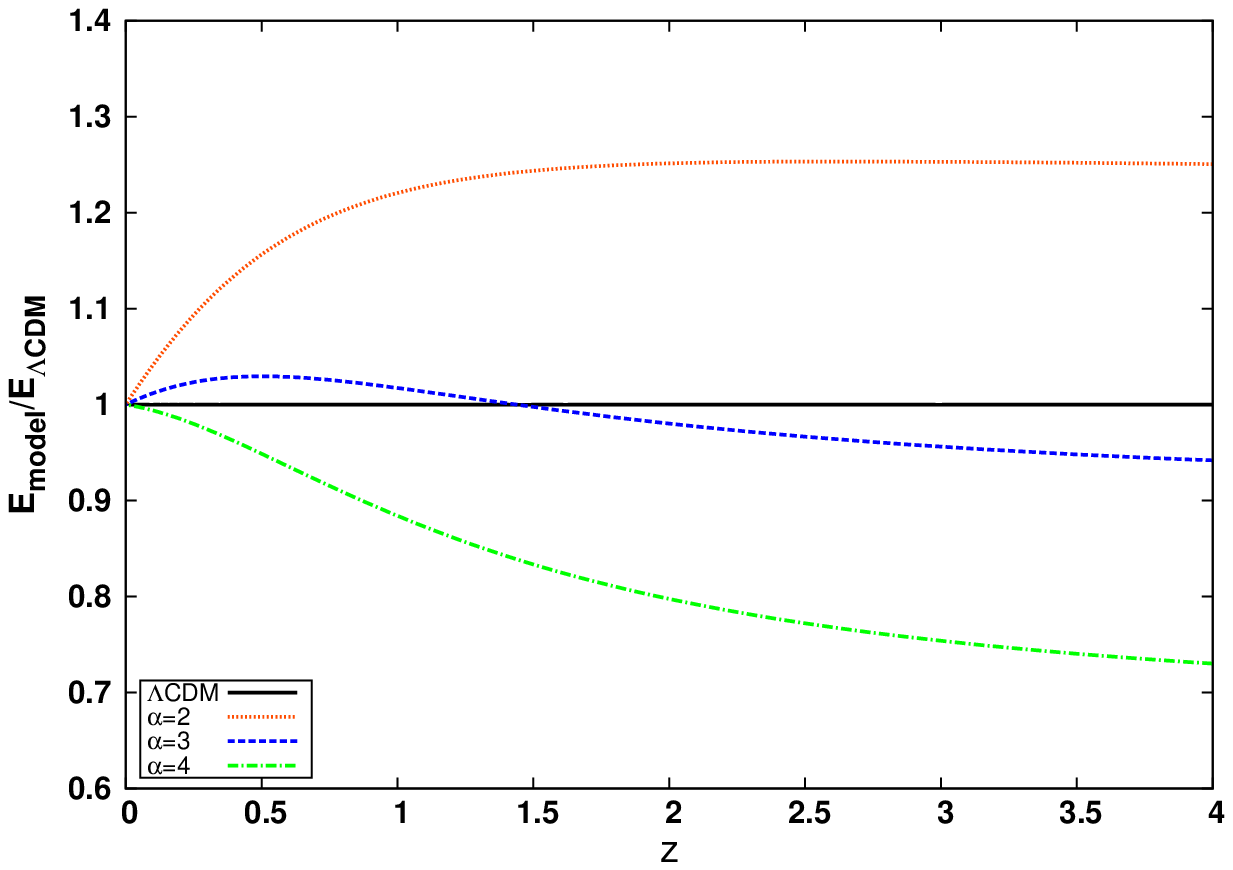}
	\includegraphics[width=0.5\textwidth]{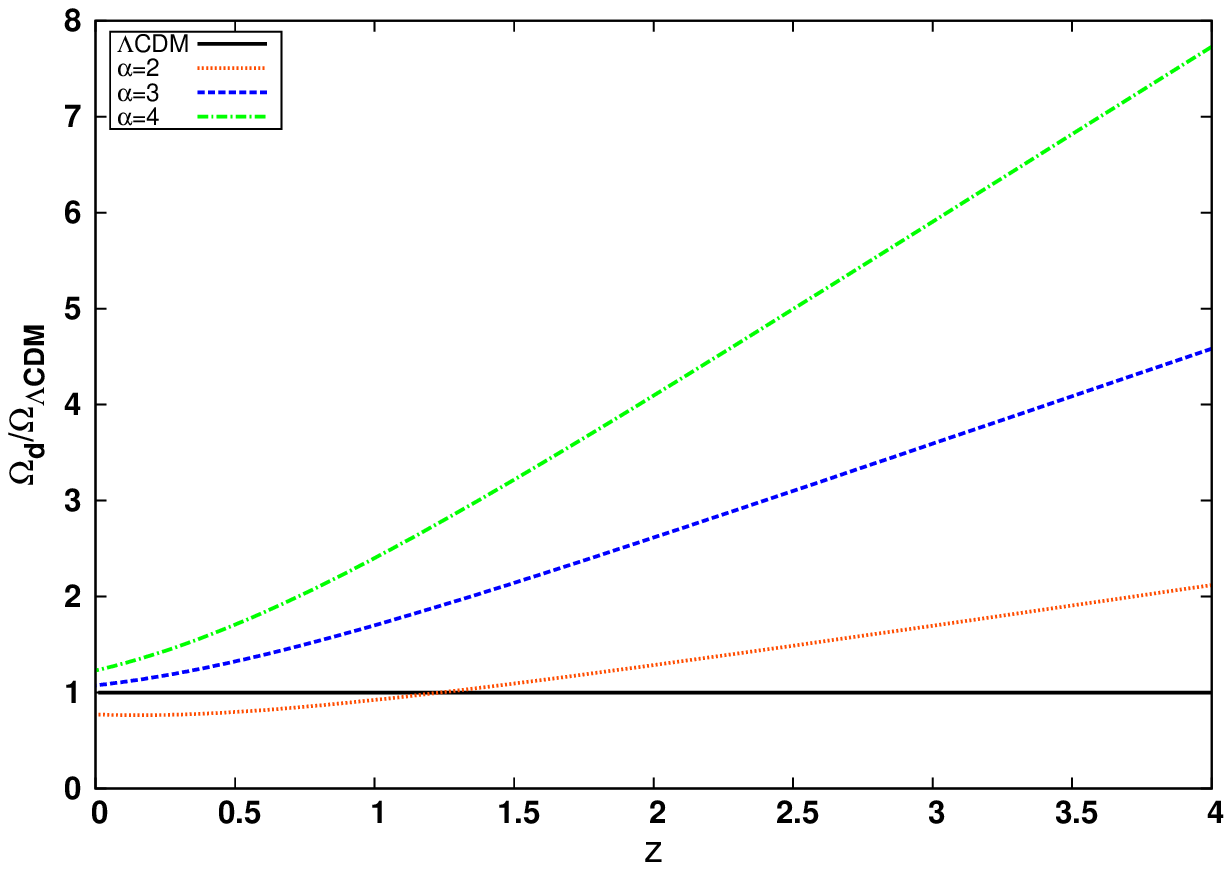}
	\caption{ The redshift evolution of the equation of
state parameter of NADE $w_{\rm d}(z)$ (top panel), ratio of dimensionless Hubble parameter of NADE model to the $\Lambda$CDM model (middle panel) and ratio of DE density parameter $\Omega_{\rm d}(z)$ to $\Omega_{\rm \Lambda}(z)$ (bottom panel) for different values of model parameter $\alpha$ considered in this work. The red dotted, blue dashed and green dot-dashed curves
correspond to NADE models with $\alpha=2$, $\alpha=3$ and $\alpha=4$, respectively. The reference $\Lambda$CDM model is shown by the black solid line.}
	\label{fig:back}
\end{figure}

Although the ADE scenario solves the casuality problem \citep{Cai2007}, it faces some problems toward describing the matter-dominated epoch \citep{Neupane:2007fw,Wei:2007ty, Wei:2007xu}.To solve these new problems, the authors in\cite{Wei:2007ty} proposed a new version of ADE dubbed the new agegraphic dark energy (NADE) model, in which they use conformal time $\eta$ instead of cosmic time $T$. The energy density in the NADE model is given by \citep{Wei:2007ty}
\begin{equation}\label{NADE}
\rho_{\rm d}=\dfrac{3 \alpha^2 m_{\rm p}^2}{{\eta}^2} \;,
\end{equation}

where $ \eta = \int_{0}^{a} \dfrac{da}{a^2 H} $ . Considering  the spatially flat Friedmann-Robertson-Walker
universe, the Friedmann equation for a universe containing radiation, pressureless
dust matter and NADE is given by
\begin{equation}\label{FRW}
H^2=\dfrac{1}{3  m_{\rm p}^2} (\rho_{\rm r}+\rho_{\rm m}+\rho_{\rm d}) \;,
\end{equation}
where $\rho_{\rm r}$, $\rho_{\rm m}$ and $\rho_{\rm d}$ are  energy densities of radiation, pressureless matter
and DE, respectively. Now utilizing the Friedmann equation (\ref{FRW}) and continuity
equations, respectively for radiation, pressureless matter, and DE,
\begin{eqnarray}\label{peim}
&\dot{\rho_{\rm r}}+4H \rho_{\rm r}=0\;,\\
&\dot{\rho_{\rm m}}+3H \rho_{\rm m}=0\;,\\
&\dot{\rho_{\rm d}}+3H(1+w_{\rm d}) \rho_{\rm d}=0\;,
\end{eqnarray}
the dimensionless Hubble parameter becomes
\begin{equation}\label{ee}
E(a)=\dfrac{H(a)}{H_0}=\dfrac{\Omega_{\rm m 0}a^{-3}+\Omega_{\rm r 0}a^{-4}}{1-{\Omega}_{\rm d}(a)}\;,
\end{equation}
where ${\Omega}_{\rm d}=\dfrac{\rho_{\rm d}}{3 {m_{\rm p}}^2 H^2 }$ is  the density parameter for DE and $\Omega_{\rm m 0}$ and $\Omega_{\rm r 0}$ are the present values of matter and radiation density parameters respectively. Replacing $\rho_{\rm d}$ from Eq. (\ref{NADE}) in ${\Omega}_{\rm d}$ it becomes:
\begin{equation}\label{omd}
{\Omega}_{\rm d}=\dfrac{\alpha^2}{ {\eta}^2 H^2 }\;.
\end{equation}
Differentiating with respect to cosmic time in Eq. (\ref{NADE}) and using   Eqs. (\ref{peim} and \ref{omd}) the equation of state parameter for NADE takes the form \citep[see also][]{Wei:2007ty}:
\begin{equation}\label{wd}
w_{\rm d}=-1 +\dfrac{2}{3\alpha a} \sqrt{\Omega_{\rm d}}\;.
\end{equation}
The evolution of energy density of DE in the NADE model is given by the following differential equation \citep[see also][]{Wei:2007ty}:
 \begin{equation}\label{dif}
\dfrac{d\Omega_{\rm d}}{da}=\dfrac{\Omega_{\rm d}}{a}(1-\Omega_{\rm d})(3-\dfrac{2}{\alpha a} \sqrt{\Omega_{\rm d}} )\;.
\end{equation}
Now by solving the system of coupled equations (\ref{ee}),(\ref{wd}) and (\ref{dif}), we can obtain the evolution of $\Omega_{\rm d}(a)$,$w_{\rm d}(a)$,  and $E(a)$. We solve coupled equations from the scale factor $a_i=0.0005$  which is deep enough in the matter-dominated epoch \citep[see also][]{ Wei:2007xu}.  Hence the initial conditions can be chosen at the matter-dominated epoch where $H^{2}\propto \rho_{\rm m}\propto a^{−3}$, and $\eta \propto a^{1/2}$. Using Eqs. (\ref{NADE}) and (\ref{omd}) we find $\Omega_{\rm d}(a_{\rm i})\simeq\dfrac{\alpha^2 {a_{\rm i}}^2}{4}$
and using Eq.(\ref{wd}) we have $w_{\rm d}(a_{\rm i})\simeq-2/3$ \citep{ Wei:2007xu}. In Fig.\ref{fig:back} we show the evolution of background quantities in NADE cosmology, equation of
state parameter of NADE $w_{\rm d}(z)$(top panel), the ratio of dimensionless Hubble parameter $E(z)$ of the NADE model to that of the $\Lambda$CDM model $E_{\rm \Lambda}$ (middle panel) and the ratio of DE density parameter $\Omega_{\rm d}(z)$ to that of the $\Lambda$CDM model $\Omega_{\rm \Lambda}(z)$ (bottom panel), for different values of model parameter $\alpha$ considered in this work. The red-dotted, blue-dashed, and green dot-dashed curves
correspond to NADE models with $\alpha =2$, $\alpha=3$ and $\alpha=4$, respectively. Also the concordance $\Lambda$CDM model is shown by the black solid line. As we can see in the top panel of Fig.\ref{fig:back} for all selected values of $\alpha$, the NADE EoS parameter obeys the inequality $ -1< w_{\rm d} <-2/3$ and thus it cannot enter in the phantom regime at all. Also, we see that by increasing the value of $\alpha$,  $w_{\rm d}$ decreases. The middle panel shows the evolution of dimensionless Hubble parameter $E(z)$. We see that $E$ for $\alpha=2$ ($\alpha=4$) is higher (smaller) than that of the $\Lambda$CDM model throughout
its history. For the case $\alpha=3$, $E$ is higher than $\Lambda$CDM at low redshifts but at higher redshifts  it falls down and becomes smaller than $\Lambda$CDM model. In analogy with the behavior of  $w_{\rm d}$, by increasing the value of $\alpha$,  $E(z)$ decreases. In the bottom panel we see that  DE density parameter $\Omega_{\rm d}$ increases with $\alpha$, as we expect from Eq.(\ref{NADE}). For all values of $\alpha$, at high redshifts, $\Omega_{\rm d}$ reduces as expected at the matter-dominated epoch. But since the decreasing of the $\Omega_{\rm \Lambda}$ is faster than $\Omega_{\rm d}$ of NADE, therefore, at high redshifts the ratio of $\Omega_{\rm d}/\Omega_{\rm \Lambda}$ increases for all values of $\alpha$.

\section{SPHERICAL COLLAPSE IN NADE COSMOLOGIES }\label{growth}

In this section we extend the SCM in the framework of NADE cosmologies. For this purpose, we first review the basic equations
used to obtain the characteristic parameters of SCM in  NADE cosmologies. In the scenario of structure formation, several attempts have
been made to derive the differential equations governing the evolution of matter and DE perturbations. Some of these  attempts have
been made to investigate the equations in a matter-dominated universe  \citep{Bernardeau1994,Padmanabhan1996,Ohta2003,Ohta2004}. In the work of \cite{Abramo2007}, the equation for the evolution of $\delta_{\rm m}$ was generalized to a universe containing a dynamical DE component.

The important note regarding the perturbations of DE is related to adiabatic sound speed. The authors of \citep{Kim:2007iv} showed that the adiabatic sound speed of agegraphic DE models is imaginary. In fact, the adiabatic sound speed of most DE models such as quintessence DE models with constant $w_{\rm de}$ is imaginary which causes the unphysical instability of DE perturbations. To overcome this problem, we can consider the perturbations of entropy. In the presence of entropy perturbation, one can define the effective sound speed $c_{\rm eff}$ for DE which is basically null or positive. In the linear regime ($\delta\ll 1$), the cosmological observations favor a small effective sound speed $c^2_{\rm eff}\leq 0.001$ for DE  (the speed of light $c=1$) \citep{Xu:2012zm,Mehrabi:2015hva}. In particular, performing the MCMC statistical analysis,  authors of  \citep{Mehrabi:2015hva}showed that the peak of the likelihood function happens at $c_{\rm eff}=0$. However in the nonlinear regime ($\delta>1$), which will be important in SCM, $c_{\rm eff}$ is a free parameter in the range of $[0,1]$. In this work, we consider two extreme  cases: $c_{\rm eff}=0$ and $c_{\rm eff}=1$  based on the following arguments. In the case of $c_{\rm eff}=1$ ( homogeneous DE) the Jeans length of DE is equal to or larger than the Hubble length and consequently the DE perturbations inside the Hubble horizon cannot grow. In this case DE distributes uniformly and only matter perturbation grows to form cosmic structures. In fact the DE component affects the perturbations of matter through changing the Hubble expansion in background cosmology. On the other hand,  the limiting case of $c_{\rm eff}=0$ ( clustered DE) results in the null value for the Jeans length scale of DE (similar to pressureless  matter). In this case the perturbations of DE can grow due to gravitational instability similar to matter perturbations \citep[see also][]{Batista:2013oca}. Notice that because of negative pressure, the amplitude of DE perturbations is much smaller than the amplitude of matter perturbations. Another important issue is that assuming $c_{\rm eff}=0$ causes the comoving collapse of DE and dark matter perturbations, therefore equations for the evolution of SCM are easily simplified.
The  equations for the evolution of  matter and dark energy perturbations ($\delta_{\rm m}$ and $\delta_{\rm d}$) in SCM (without the contribution of shear and rotation) are given by\citep{Pace:2014taa} 

\begin{equation}\label{nl1}
\acute{\delta_{\rm m}}+(1+\delta_{\rm m})\dfrac{\tilde{\theta}}{a}=0\;,
\end{equation}
\begin{equation}\label{nl2}
\acute{\delta_{\rm d}}-\dfrac{3}{a}w_{\rm d}\delta_{\rm d}+(1+w_{\rm d}+\delta_{\rm d})\dfrac{\tilde{\theta}}{a}=0\;,
\end{equation}
\begin{eqnarray} \label{nl3}
\tilde{\theta}^{\prime}+(\frac{2}{a}+\frac{E^{\prime}}{E})\tilde{\theta}+\frac{{\tilde{\theta}}^2}{3a}+\frac{3}{2a}(\Omega_{\rm m}\delta_{\rm m}+\Omega_{\rm d}\delta_{\rm d})=0\;,
\end{eqnarray}
where $\tilde{\theta}=\dfrac{\theta}{H}$ is the dimensionless divergence of the comoving
peculiar velocity for both nonrelativistic matter and DE.
The linearized Equations. (\ref{nl1}), (\ref{nl2}) and (\ref{nl3}) read

\begin{equation}\label{l1}
\acute{\delta_{\rm m}}+\dfrac{\tilde{\theta}}{a}=0\;,
\end{equation}
\begin{equation}\label{l2}
\acute{\delta_{\rm d}}-\dfrac{3}{a}w_{\rm d}\delta_{\rm d}+(1+w_{\rm d})\dfrac{\tilde{\theta}}{a}=0\;,
\end{equation}
\begin{equation}\label{l3}
\acute{\tilde{\theta}}+(\dfrac{2}{a}+\dfrac{E^{\prime}}{E})\tilde{\theta}+\dfrac{3}{2a}(\Omega_{\rm m}\delta_{\rm m}+\Omega_{\rm d}\delta_{\rm d})=0\;.
\end{equation}
For appropriate initial conditions, we will obtain the linear
overdensity  $\delta_{\rm m}$ for nonrelativistic matter  and $\delta_{\rm d}$ for DE at any redshift $z$. These equations are
also used to determine the time evolution of the growth factor if suitable initial conditions are used. To determine the appropriate initial conditions, we start by considering nonlinear equations (\ref{nl1}), (\ref{nl2}) and (\ref{nl3}). Since at collapse
time $a_{\rm c}$ the collapsing sphere falls to the center, its overdensity $\delta_{\rm m}$ basically
becomes infinite. Thus, we search for an initial
matter density contrast $\delta_{\rm mi}$ such that the $\delta_{\rm m}$  from solving the nonlinear equations diverges (numerically, we assume this to be achieved when $\delta_{\rm m}\geq{10}^7$)
at the chosen collapse time. Once $\delta_{\rm mi}$ is found, we use this value as one of the initial conditions in our linear differential equations (\ref{l1}), (\ref{l2}) and (\ref{l3}) to find the linear threshold parameter $\delta_{\rm c}$ as one of the main quantities in SCM scenario. In fact in the context of the SCM when $\delta_{\rm m}^{\rm linear}\geq\delta_{\rm c}$ the corresponding perturbed region is virialized. Since we are dealing with three differential equations, three initial
conditions have to be chosen. Two others are initial
values for the DE overdensity $\delta_{\rm di}$ and peculiar velocity perturbation ${\tilde{\theta}}_{\rm i}$, where both of them are related to $\delta_{\rm mi}$, via \citep{Batista:2013oca,Pace:2014taa}
\begin{equation}\label{in1}
\delta_{\rm di}=\dfrac{\mu}{\mu-3w_{\rm d}}(1+w_{\rm di})\delta_{\rm mi}\;,
\end{equation}
\begin{equation}\label{in2}
{\tilde{\theta}}_{\rm i}=-\mu\delta_{\rm mi}\;.
\end{equation}

In the case of an Einstein de Sitter (EdS) universe we have $\mu=1$. However, in DE cosmologies it has been shown that there is a small deviation from unity
\citep{Batista:2013oca}. Since at high redshifts the contribution of DE is negligible, we approximately set $\mu=1$ in Eqs. (\ref{in1}) and (\ref{in2}) to obtain the two remaining initial conditions for solving coupled linear equations (\ref{l1}), (\ref{l2}) and (\ref{l3}). In homogeneous DE with ${c_{\rm eff}}=1$, we have $\delta_{\rm d}=0$ and the systems of Eqs. (\ref{nl1})-(\ref{nl3}) and (\ref{l1})-(\ref{l3}) are respectively reduced to Eqs. (18) and (19) in \citep{Pace2010} as expected.

\subsection{Growth factor and ISW effect}
\begin{figure} 
	\centering
	\includegraphics[width=0.5\textwidth]{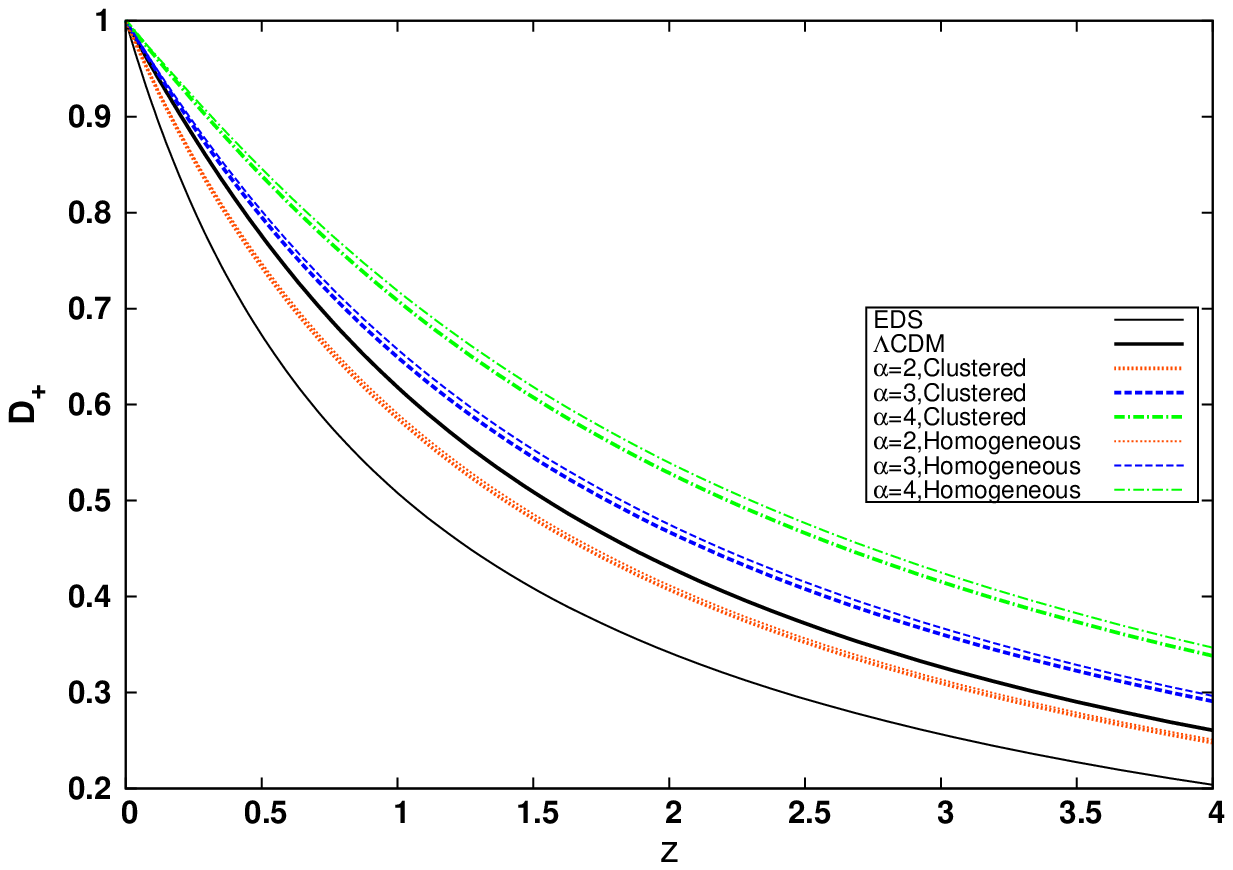}
    \includegraphics[width=0.5\textwidth]{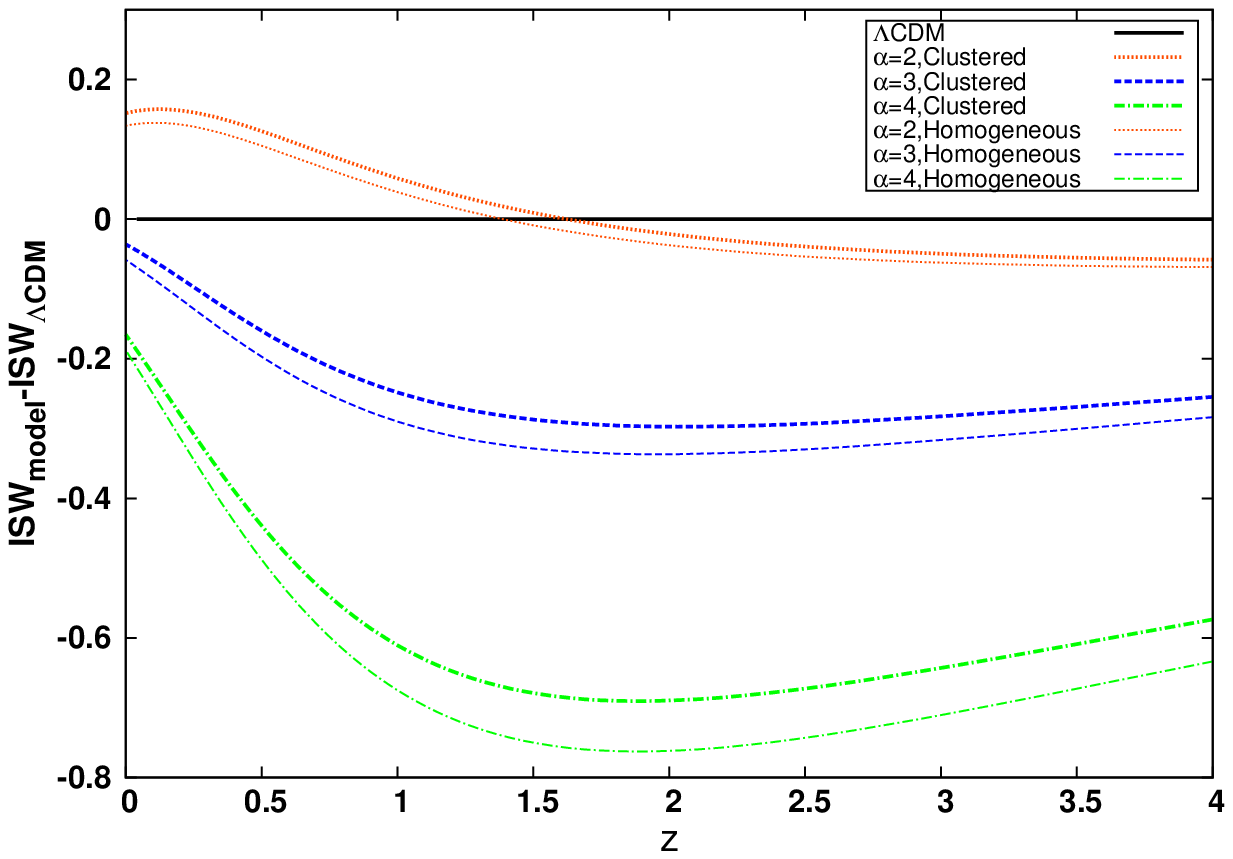}
	\caption{ The evolution of growth factor (top panel) and ISW (bottom panel) as a function of redshift $z$  for different values of model parameter $\alpha$ considered in this work. The red dotted, blue dashed and green dot-dashed curves
correspond to NADE models with $\alpha=2$, $\alpha=3$ and $\alpha=4$, respectively. Thick and thin curves represent clustered and homogeneous NADE respectively. 
The reference $\Lambda$CDM (EdS) model is shown by the thick (thin) solid black line.}
	\label{fig:d}
\end{figure}

Here we follow the linear growth of perturbations of nonrelativistic
dust matter by solving coupled linear equations (\ref{l1}), (\ref{l2}) and (\ref{l3}). We compute the linear growth factor $D_+(a) =\delta_{\rm m}(a)/\delta_{\rm m}(a = 1)$  \citep[for a similar discussion, see also][]{Copeland2006, Nesseris:2007pa,Tsujikawa2008, Pettorino:2008ez,
Basilakos:2009mz, Lee2011,Rezaei:2017yyj}. Figure \ref{fig:d} (top panel) shows the variations of
the growth factor as
a function of redshift $z$ for different values of model parameter $\alpha$. The growth factor of perturbations in NADE models with  selected model parameter $\alpha=3,4$ ($\alpha=2$) in this analysis is larger (smaller) than the $\Lambda$CDM universe. All NADE models and concordance $\Lambda$CDM models result in a bigger growth factor than the EdS universe. This result is expected since in former models, DE suppresses the growth of matter perturbations and in EdS universe this suppression does not exist. Therefore in DE models, the initial matter perturbations should grow with a larger growth factor than the EdS universe to exhibit the large scale structures observed today.
Also, for larger values of model parameter $\alpha$, the energy density of NADE becomes more significant as we expect from Eq.(\ref{NADE}) so that the suppression process of matter perturbations is enhanced. Thus for the case $\alpha=4$, we predict the largest value for the growth factor.
Moreover, for all values of $\alpha$, growth factor in homogeneous NADE cases is bigger than those obtained in clustered  NADE cases respectively. In fact when DE can cluster, the amount of clustered DE behaves as DM and amplifies the formation of cosmic structures. 
The study of the growth factor is important also for the evaluation
of the integrated Sachs-Wolfe (ISW) \citep{Sachs:1967er}. The ISW effect can distinguish the cosmological
constant from other models of dark energy \citep{Cooray:2001ab,Dent:2008ek}. The
ISW effect is due to the interaction of CMB photons with a time
varying gravitational potential. The relative change of the CMB
temperature is given by

\begin{equation}\label{is}
\tau=\dfrac{\Delta T}{T_{\rm CMB}}=\dfrac{2}{c^3}{\int_0}^{\chi_{\rm H}}d\chi a^2H(a)\dfrac{\partial}{\partial a}(\Phi-\Psi)\;,
\end{equation}

where $\chi_{\rm H}$ is the horizon distance. The gravitational potentials are
related via the Poisson equation to the matter overdensity. The ISW effect is therefore
proportional to the quantity $dD_+(a)/da$. Dark energy perturbations affect the low $l$ quadrupole
in the CMB angular power spectrum through the ISW
effect\citep{Weller:2003hw,Bean:2003fb}. Here we are in particular, interested in the late ISW
effect because it is affected by the dark energy component.The ISW
effect depends on the time derivative of the gravitational potential
$\Phi$ and the overdensity $\delta$ via the Poisson equation \citep{Pace:2013pea} .
In the bottom panel of Fig.\ref{fig:d} we present the difference between ISW effect of the NADE model and that obtained in $\Lambda$CDM. For all values of NADE model parameter $\alpha$, since dark
energy perturbations affect the matter perturbations,  the value of ISW for clustered NADE is always closer to the predictions in $\Lambda$CDM, compare to the results of the homogeneous NADE. The differences from the $\Lambda$CDM model becomes smaller at low redshifts due to the fact that
the NADE equation of state becomes closer to
$w=-1$.

\subsection{Parameters of the SCM}
Now we calculate two main quantities of SCM, the linear overdensity parameter $\delta_{\rm c}$ and the virial overdensity parameter $\Delta_{\rm vir}$ in the context of NADE cosmologies. The quantity $\delta_{\rm c}$ together
with the linear growth factor $D_+(z)$ are used to calculate the mass function of virialized halos \citep[see e.g.][]{Press1974,Sheth:1999su,Sheth2002}. 
To calculate $\delta_{\rm c}$ in NADE cosmologies, we use the following fitting function obtained by \cite{Kitayama:1996ne,Weinberg:2002rd} 
\begin{equation}\label{beta}
\delta_{\rm c}(z_{\rm c})=\dfrac{3(12\pi)^{2/3}}{20}(1+\beta \log \Omega_{\rm m}(z))\;.
\end{equation}
Different values of model parameter $\alpha$, result in different slope parameters $\beta$ presented in Table\ref{tab:beta}  for homogeneous and clustered NADE models, respectively.

\begin{table}
 \centering
 \caption{The results for fitting parameter $\beta$ in Eq.\ref{beta}.}
\begin{tabular}{c  c  c c}
\hline \hline
 Model  & $\alpha=2$ & $\alpha=3$ & $\alpha=4$\\
 \hline
Homogeneous DE & 0.00469021 &0.00571213 & 0.00602396\\
 \hline 
 Clustered DE & 0.00477487 & 0.00557702 &0.00577207\\

 \hline \hline
\end{tabular}\label{tab:beta}
\end{table}

\begin{figure} 
	\centering
	\includegraphics[width=0.5\textwidth]{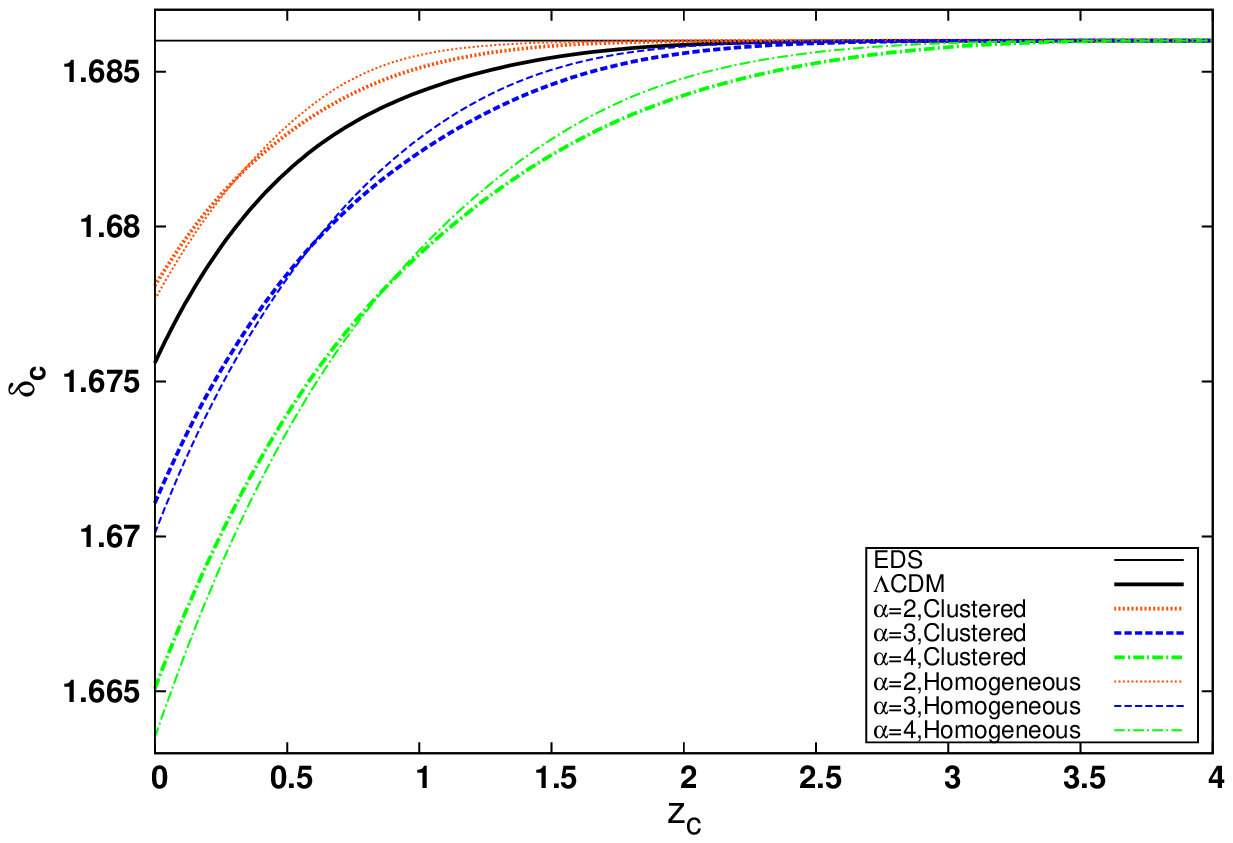}
    \includegraphics[width=0.5\textwidth]{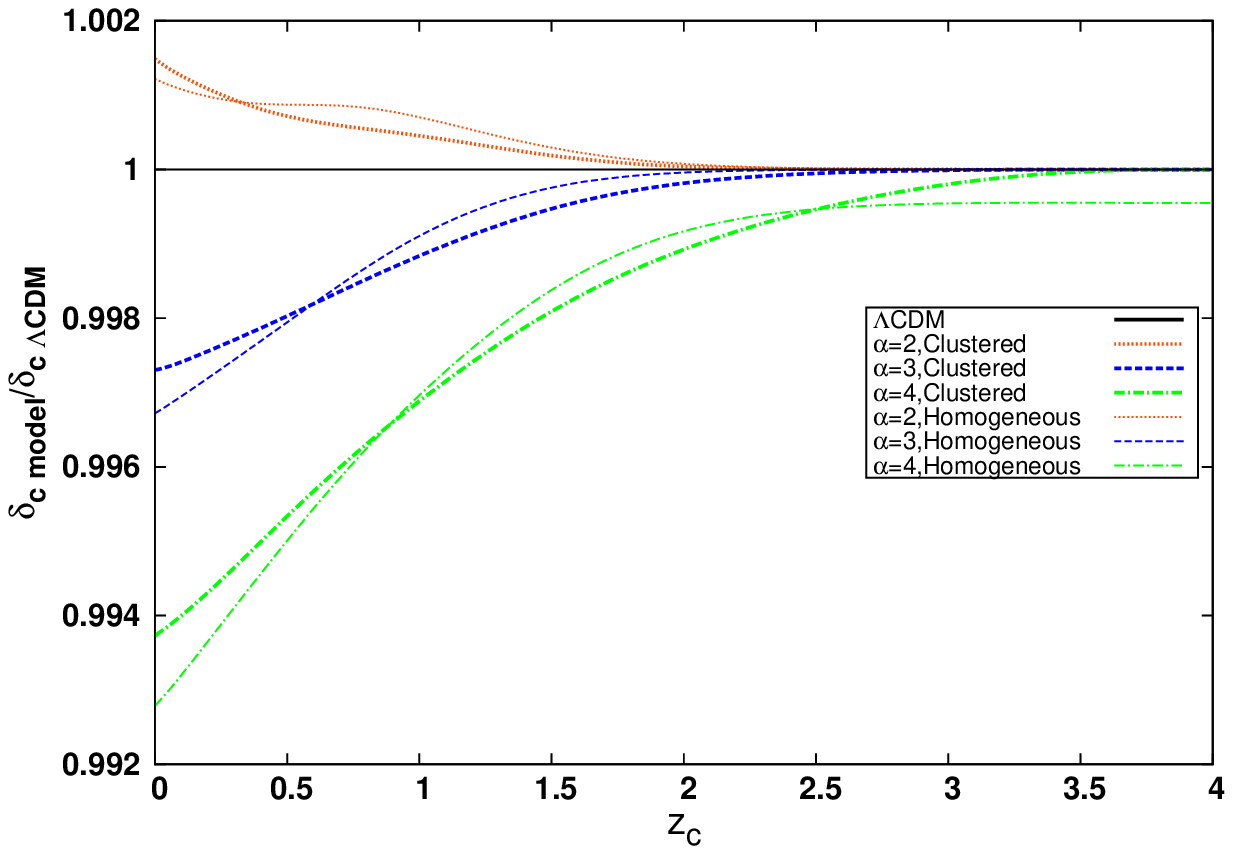}
	\caption{ Linear overdensity parameter
$\delta_{\rm c}$ as a function of $z_{\rm c}$(top panel) and ratio of linear overdensity parameter
of NADE  to that of the $\Lambda$CDM as a function of $z_{\rm c}$(bottom panel), for different values of model parameter $\alpha$ considered in this work. Line styles and colors are the same as in Fig.\ref{fig:d}.}
	\label{fig:dc}
\end{figure}

\begin{figure} 
	\centering
	\includegraphics[width=0.5\textwidth]{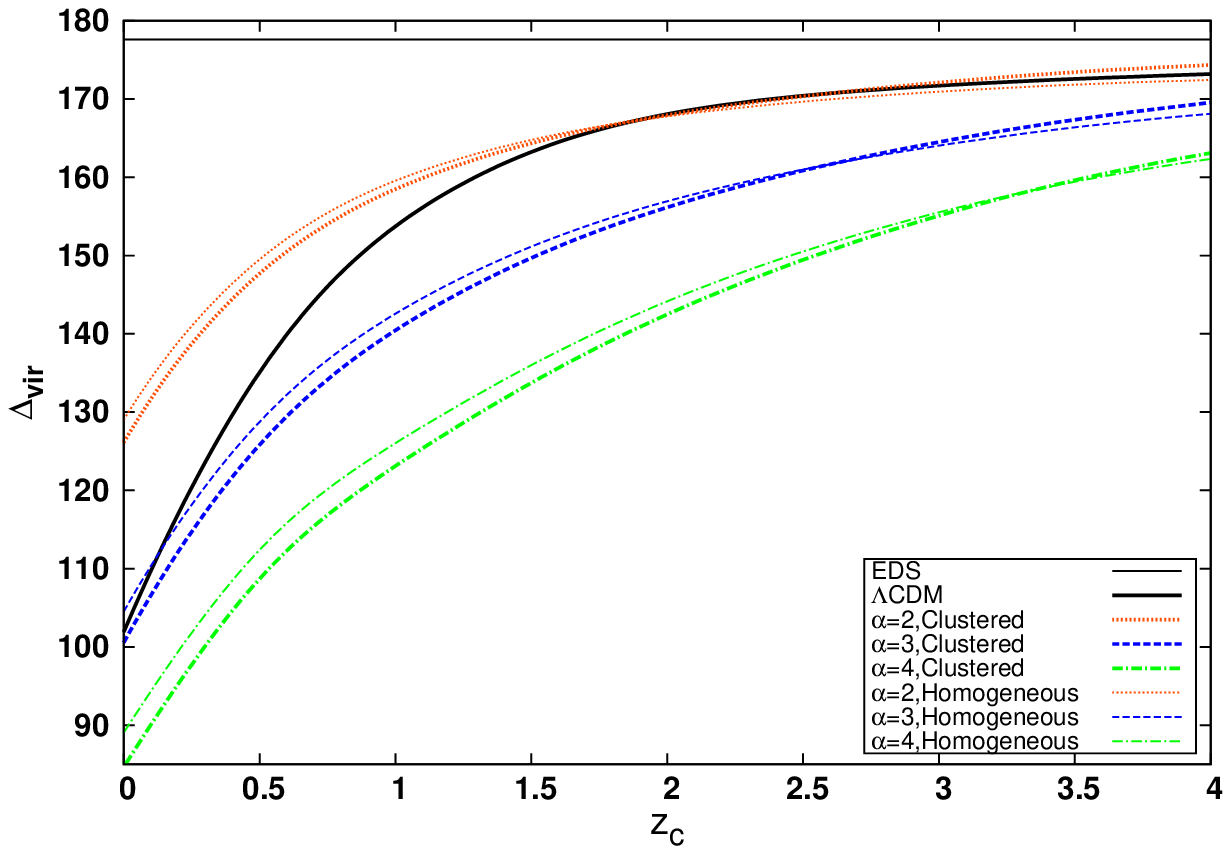}
      \includegraphics[width=0.5\textwidth]{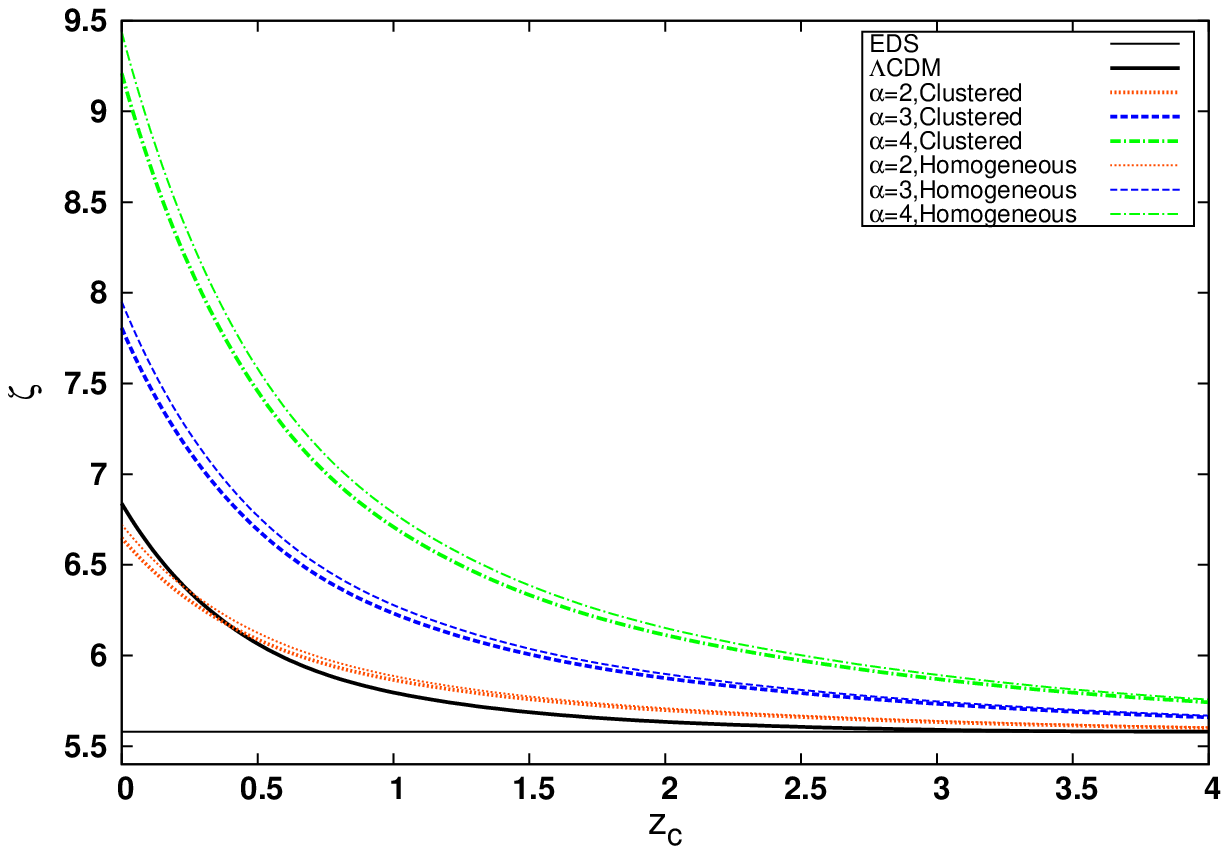}
	\caption{Redshift evolution of virial overdensity
parameter$\Delta_{\rm vir}$(top panel) and  turnaround overdensity $\zeta$ (bottom panel) for different values of model parameter $\alpha$ considered in this work. Line styles and colors are the same as in Fig.\ref{fig:d}.} \label{fig:dv}
\end{figure}

The other parameter in SCM is the virial overdensity $\Delta_{\rm vir}$. The virial overdensity is used to define the size of halos. This quantity is given by $\Delta_{\rm vir} = \delta_{\rm nl} + 1 = \zeta(x/y)^3$ 
where $x = a/a_{\rm t}$ is the normalized
scale factor and $y$ is the radius of the sphere normalized to its value
at the turnaround and $\zeta$ is the overdensity at the turnaround epoch \citep[see also][]{Pace2010}. Our results for the evolution of $\delta_{\rm c}$, $\Delta_{\rm vir}$ and $\zeta$ are presented in Figs. (\ref{fig:dc}) and (\ref{fig:dv}).
In Fig.\ref{fig:dc} we show the time evolution of the linear overdensity parameter $\delta_{\rm c}$ in the NADE model (top panel) and the ratio of the linear overdensity parameter of NADE to that of $\Lambda$CDM (bottom panel).
We see that the NADE models with $\alpha=3$ and $\alpha=4$ ( $\alpha=2$) always have a lower (higher) $\delta_{\rm c}(z)$ with respect to the $\Lambda$CDM model. We also observe that at $z_{\rm c}=0$, the $\delta_{\rm c}$ in clustered NADE models is larger compared to homogeneous cases. The difference between  $\delta_{\rm c}$ of NADE models compared to that of  $\Lambda$CDM is smaller than $0.8 \%$. NADE models, similar to  $\Lambda$CDM cosmology, asymptotically approach the EdS limit at high redshift, where we can ignore the effects of DE.

Figure \ref{fig:dv} shows the evolution of the virial overdensity
parameter $\Delta_{\rm vir}(z)$ (top panel) and  turnaround overdensity $\zeta$ (bottom panel). In all models, $\Delta_{\rm vir}$ tends to EdS value $178$ at high redshifts, as expected. At low redshifts, decrements of $\Delta_{\rm vir}$ indicate that low dense virialized halos are formed  in NADE and $\Lambda$CDM models compared to the EdS model. Particularly in the case of the NADE model with $\alpha=4$, the density of dark matter in virialized halos is $\sim 50\%$  lower than that of the EdS model. This value is roughly $44\%$ for the $\Lambda$CDM model and the NADE model with $\alpha=3$. In the case of $\alpha=2$ we observe this vale as $\sim 27\%$. The lower density of virialized halos in NADE and $\Lambda$CDM models  than the EdS universe can be interpreted as the affect of DE on the process of virialization. In fact DE prevents more collapse and consequently halos virialize at a larger radius with a lower density.
We also conclude that $\Delta_{\rm vir}$ in homogeneous NADE models is larger than clustered NADE.
Finally, the evolution of turnaround overdensity $\zeta$ is shown in the bottom panel of Fig.\ref{fig:dv}. As expected, in the limiting case of the EdS model, $\zeta=5.6$. At high redshifts, $\zeta$ tends to the EdS value $\zeta=5.6$ representing the early matter-dominated era. In both clustered
and homogeneous versions of NADE models with $\alpha=3$ and $4$,  $\zeta$ is larger than that of  the concordance $\Lambda$CDM model. Moreover, $\zeta$ for clustered NADE is smaller than the homogeneous version which shows that in homogeneous NADE, the perturbed spherical
region detaches from the Hubble flow with higher
overdensity compared to the clustered cases.

\section{Mass Function and Number of Halos}\label{sec:mass}
In this section using the Press-Schechter formalism, we compute the number of cluster-size halos 
in the context of the NADE cosmologies. In Press-Schechter formalism the abundance of
virialized halos can be expressed in terms of their mass
\citep{Press1974}. The comoving number density of virialized
halos with masses in the range of $M$ and $M+dM$ is given by
\citep{Press1974,Bond1991}
\begin{equation}\label{press}
\dfrac{dn(M,z)}{dM}=\dfrac{\rho_{\rm m0}}{M}\dfrac{d\sigma^{-1}}{dM}f(\nu)\;,
\end{equation}

where $\rho_{\rm m0}$ is the background density of matter at the present time, $\nu(M,z)=\delta_{\rm c} / \sigma$ and  $\sigma$ is the root mean square of the mass fluctuations in spheres containing the mass $M$.

 Although the standard mass function $f(\nu)=\sqrt{{2}/{\pi}} e^{-\frac{\nu}{2}}$ presented in \citep{Press1974,Bond1991} can provide
a good estimate of the predicted number density of halos, it fails
by predicting too many low-mass and too few high-mass
objects \citep{Sheth1999,Sheth2002,Lima:2004np}. Hence, in this work we use another popular fitting formula proposed by  \cite{Sheth1999,Sheth2002}
\begin{equation}\label{sheth}
f(\nu)=0.2709\sqrt{\dfrac{2}{\pi}}(1+1.1096\nu^{0.6})exp(-\dfrac{0.707 \nu^2}{2})\;.
\end{equation}
In a Gaussian density field, $\sigma$ is given by
\begin{equation}\label{sigma}
\sigma^2(R)=\dfrac{1}{2 \pi^2}{\int_0}^\infty k^2 P(k) W^2(kR) dk\;,
\end{equation}
where $R=(3M/4\pi \rho_{\rm m0})^{1/3}$ is the radius of the overdense spherical region, $W(kR)$ is the Fourier transform of a spherical top-hat profile with radius $R$ and $P(k)$ is
the linear power spectrum of density fluctuations \citep{Peebles1993}. To calculate $\sigma$, we follow the procedure presented in \citep{Abramo2007,Naderi2015}. Following \cite{Ade:2015xua}, we use the normalization of matter power spectrum $\sigma_8=0.815$ for concordance $\Lambda$CDM model. The number density of dark matter halos above a
certain  mass $M$ at collapse redshift $z$ is simply given by

\begin{equation}\label{nn}
N(\> M,z)={\int_0}^\infty \dfrac{dn(z)}{dM'}dM'\;,
\end{equation}
where we fix the above limit of integration by $M=10^{18}M_{\rm sun}h^{-1}$ as such a gigantic structure could not in practice be observed.
We now compute the predicted number density of virialized
halos for homogeneous and clustered NADE models using Eqs. (\ref{press}) and (\ref{nn}). In this case the total mass of halos is defined
by the pressureless matter perturbations. However,
it was shown that the virialization of
dark matter perturbations in the nonlinear regime depends on the
properties of DE models \citep{Lahav1991,Maor2005,Creminelli2010,Basse2011}. Thus in clustered
DE models, we should take into account the contribution of DE
perturbations to the total mass of the halos \citep{Creminelli2010,Basse2011,Batista:2013oca,Pace:2014taa}. Depending on the form of  EoS parameter, $w_{\rm d}(z)$, DE may decrease or increase the total mass of the halo.
The fraction of DE mass taken into account with respect to the mass of pressureless matter is given by:

\begin{equation}\label{epsil}
\epsilon(z)=\dfrac{m_{\rm DE}}{m_{\rm DM}}\;,
\end{equation}

where $m_{\rm DE}$ depends on what we consider as a
mass of the DE component. If we only consider the contribution of
DE perturbation, then we would have

\begin{equation}\label{mdep}
{m_{\rm DE}}^{Perturbed}=4 \pi \bar{\rho}_{\rm DE}{\int_0}^{R_{\rm vir}} dR R^2 \delta_{\rm DE}(1+3 c^2_{\rm eff})\;,
\end{equation}
but if we assume both the contributions of
DE perturbation and DE at the background level, the total mass of DE in virialized halos takes the form
\begin{equation}\label{mdet}
{m_{\rm DE}}^{Total}=4 \pi \bar{\rho}_{\rm DE}{\int_0}^{R_{\rm vir}} dR R^2 [(1+3 w_{\rm DE})+ \delta_{\rm DE}(1+3c^2_{\rm eff})]\;.
\end{equation}
Since we work in the framework of the top-hat spherical profile, the quantities inside
the collapsing region vary only with cosmic time. Thus from Eq.(\ref{mdep}) we can obtain

\begin{equation}\label{epsil1}
\epsilon(z)=\dfrac{\Omega_{\rm DE}}{\Omega_{\rm DM}}\dfrac{\delta_{\rm DE}}{1+\delta_{\rm DM}}\;,
\end{equation}
 and from Eq.(\ref{mdet}) we have

\begin{equation}\label{epsil2}
\epsilon(z)=\dfrac{\Omega_{\rm DE}}{\Omega_{\rm DM}}\dfrac{1+3 w_{\rm DE}+\delta_{\rm DE}}{1+\delta_{\rm DM}}\;.
\end{equation}

\begin{figure} 
	\centering
	\includegraphics[width=8cm]{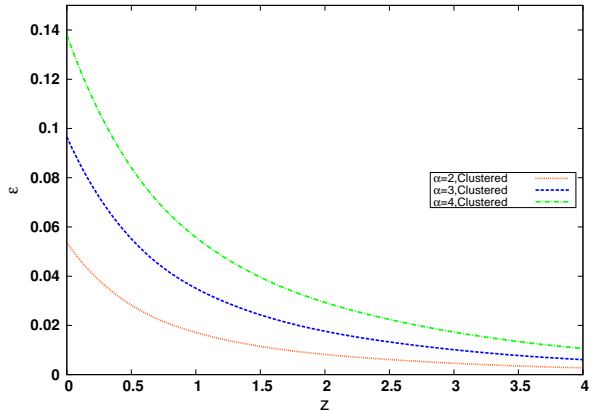}
	\caption{ The redshift evolution of the ratio of DE to dark
matter mass $\epsilon(z)$ calculated according to Eq.\ref{epsil1} for different values of NADE model parameter $\alpha$ considered in this work. The red dotted, blue dashed and green dot-dashed curves
correspond to clustered NADE model with $\alpha=2$, $\alpha=3$ and $\alpha=4$, respectively.}
	\label{fig:eps}
\end{figure}

Also the mass of dark matter is defined as \citep[see also][]{Malekjani:2015pza,Nazari-Pooya:2016bra}:

\begin{equation}\label{mdm}
{m_{\rm DM}}=4 \pi \bar{\rho}_{\rm DM}{\int_0}^{R_{\rm vir}} dR R^2 (1+ \delta_{\rm DM})\;.
\end{equation}
In this work we adopt the definition of DE mass based on Eq.(\ref{epsil1}).
 In Fig. \ref{fig:eps} we show the evolution of $\epsilon(z)$ from Eq.(\ref{epsil1}). One can see, at high redshift,
where the contribution of dark energy is less important, $\epsilon$ for all values of $\alpha$ becomes negligible. Also, for different values of model parameter $\alpha$, the amount of $\epsilon$ in clusters becomes larger by increasing the value of $\alpha$.

To compute the number density of virialized halos in clustered DE, one should assume the
presence of the DE mass correction. Following the procedure outlined
in \cite{Batista:2013oca,Pace:2014taa}, the mass of halos in clustered DE models is $M(1-\epsilon)$. Hence, the corrected mass function can be rewritten as \citep{Batista:2013oca} 
\begin{equation}\label{presscor}
\dfrac{dn(M,z)}{dM}=\dfrac{\rho_{\rm m0}}{M(1-\epsilon)}\dfrac{d \nu(M,z)}{dM}f(\nu)\;.
\end{equation}
In the case of clustered NADE models, we insert Eq.(\ref{presscor}) into Eq.(\ref{nn}) in order to calculate the number density of virialized halos.

\begin{figure*} 
	\centering
	\includegraphics[width=8cm]{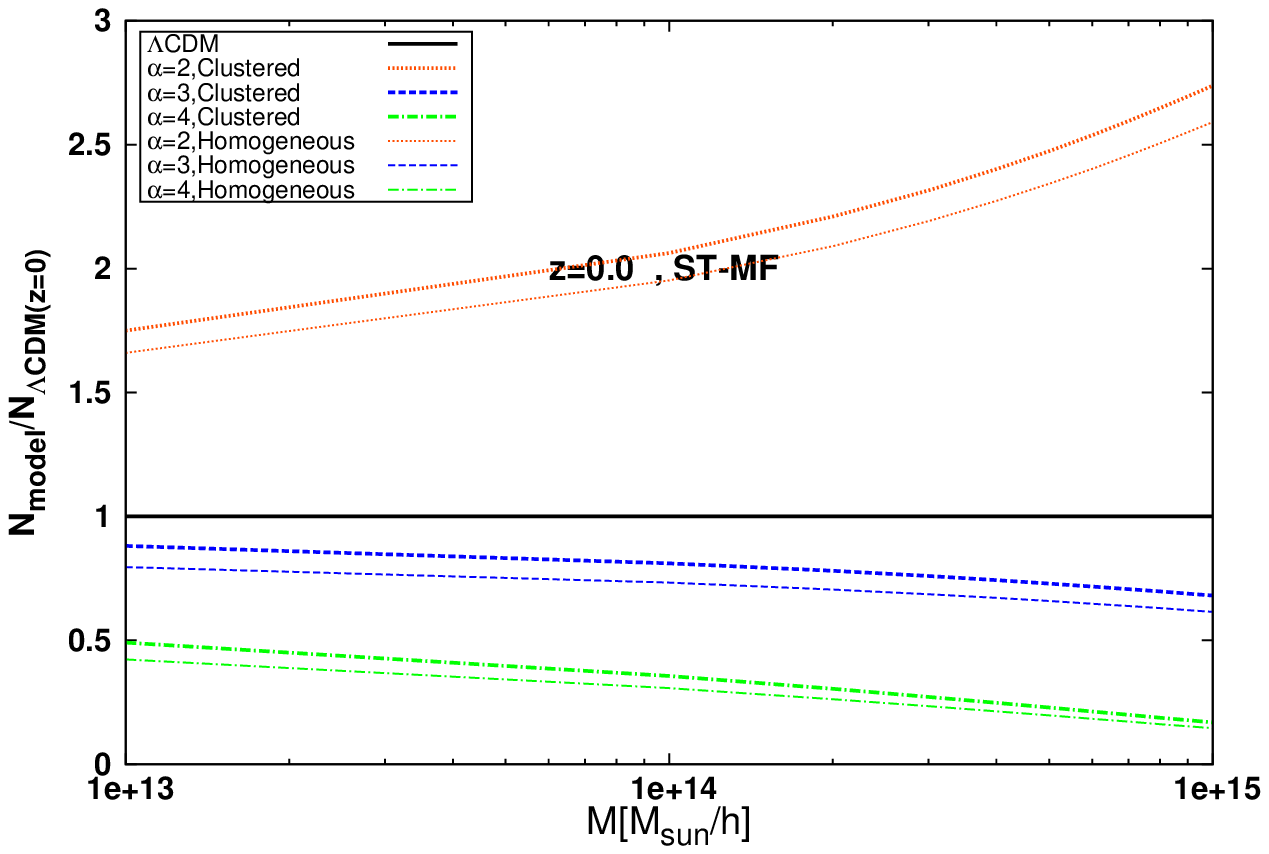}
    \includegraphics[width=8cm]{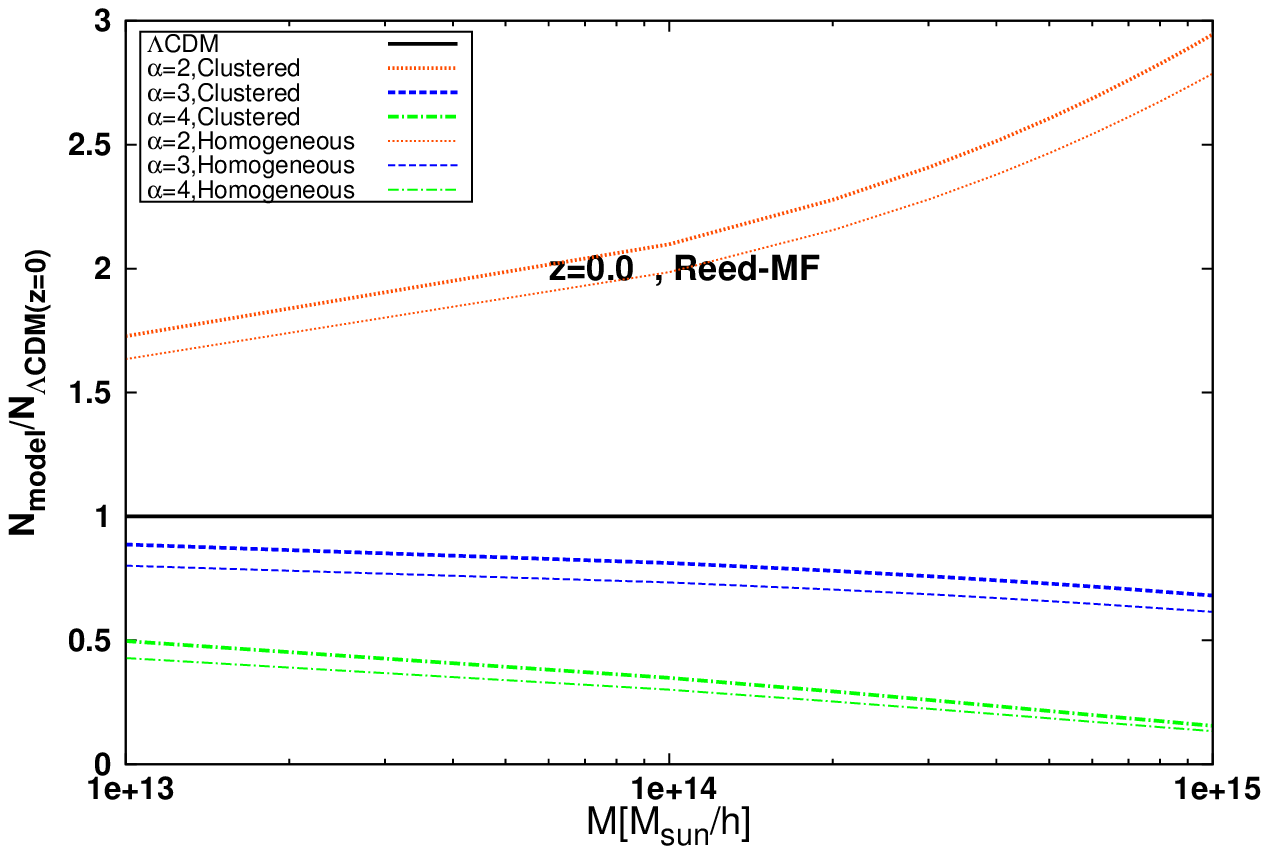}
	\includegraphics[width=8cm]{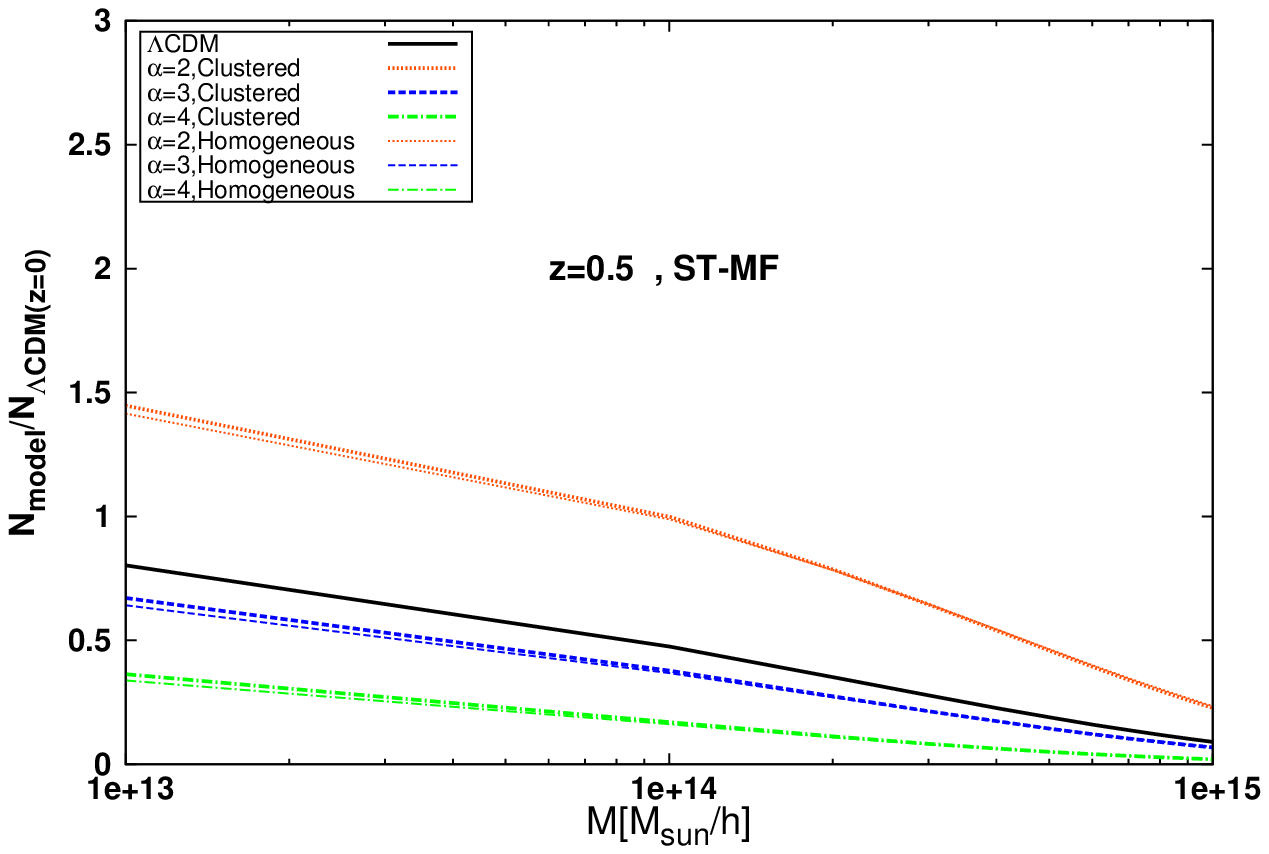}
    \includegraphics[width=8cm]{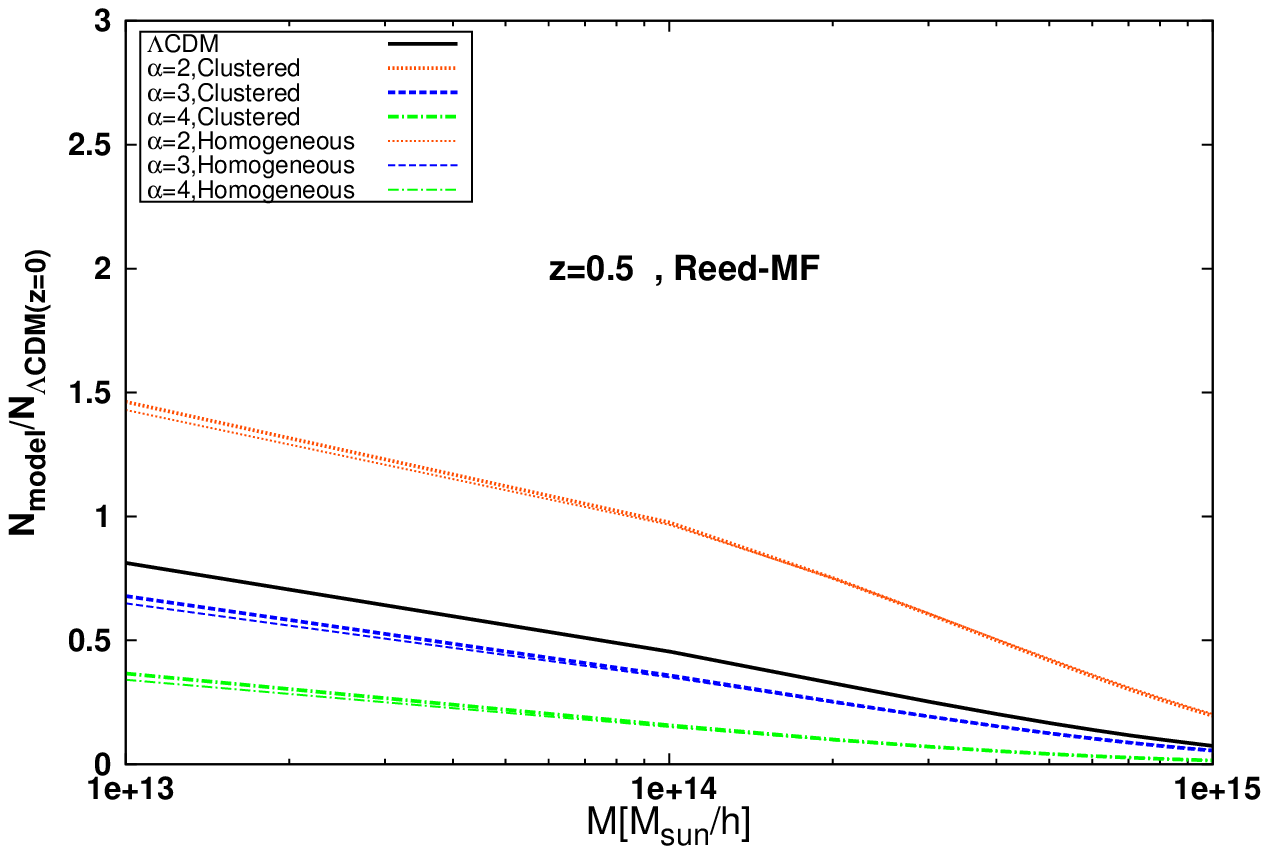}
	\includegraphics[width=8cm]{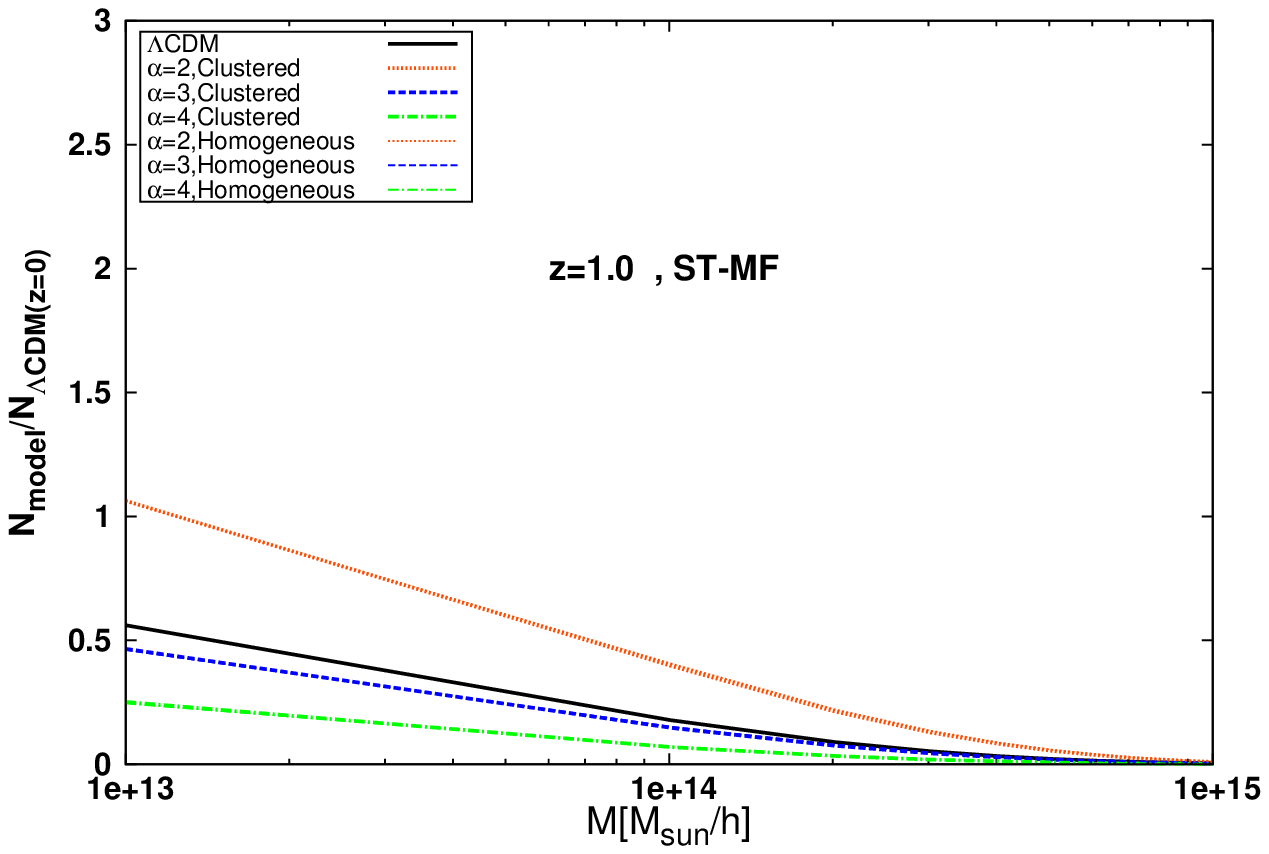}
    \includegraphics[width=8cm]{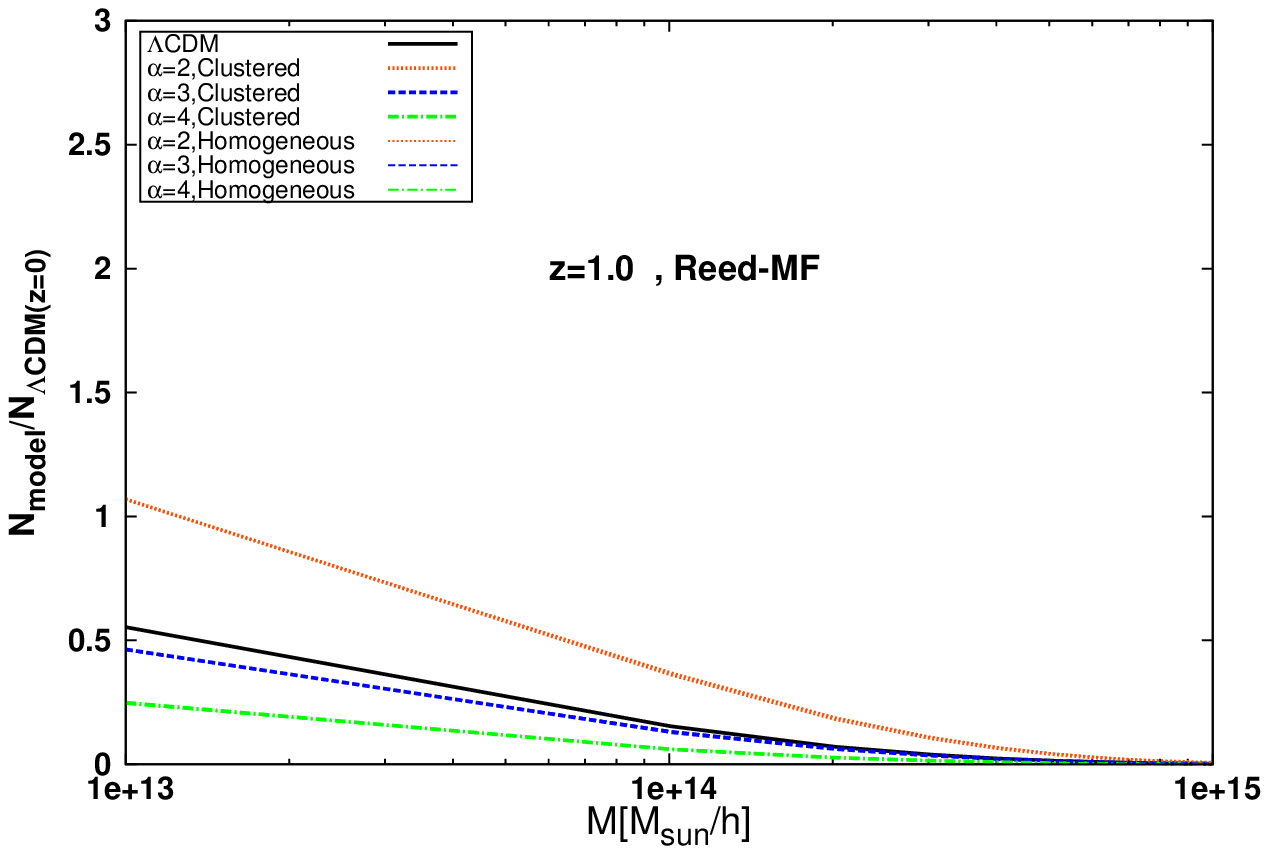}
	\includegraphics[width=8cm]{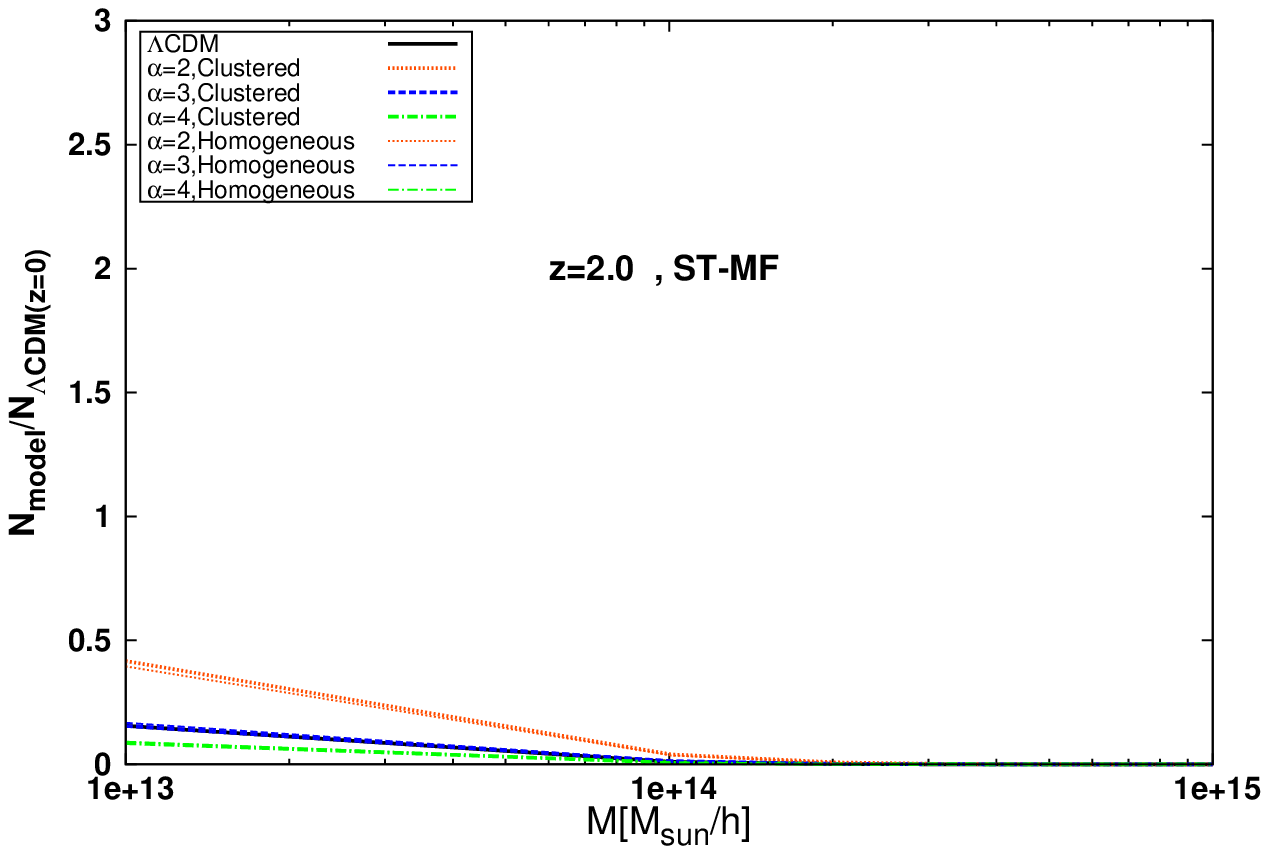}
	\includegraphics[width=8cm]{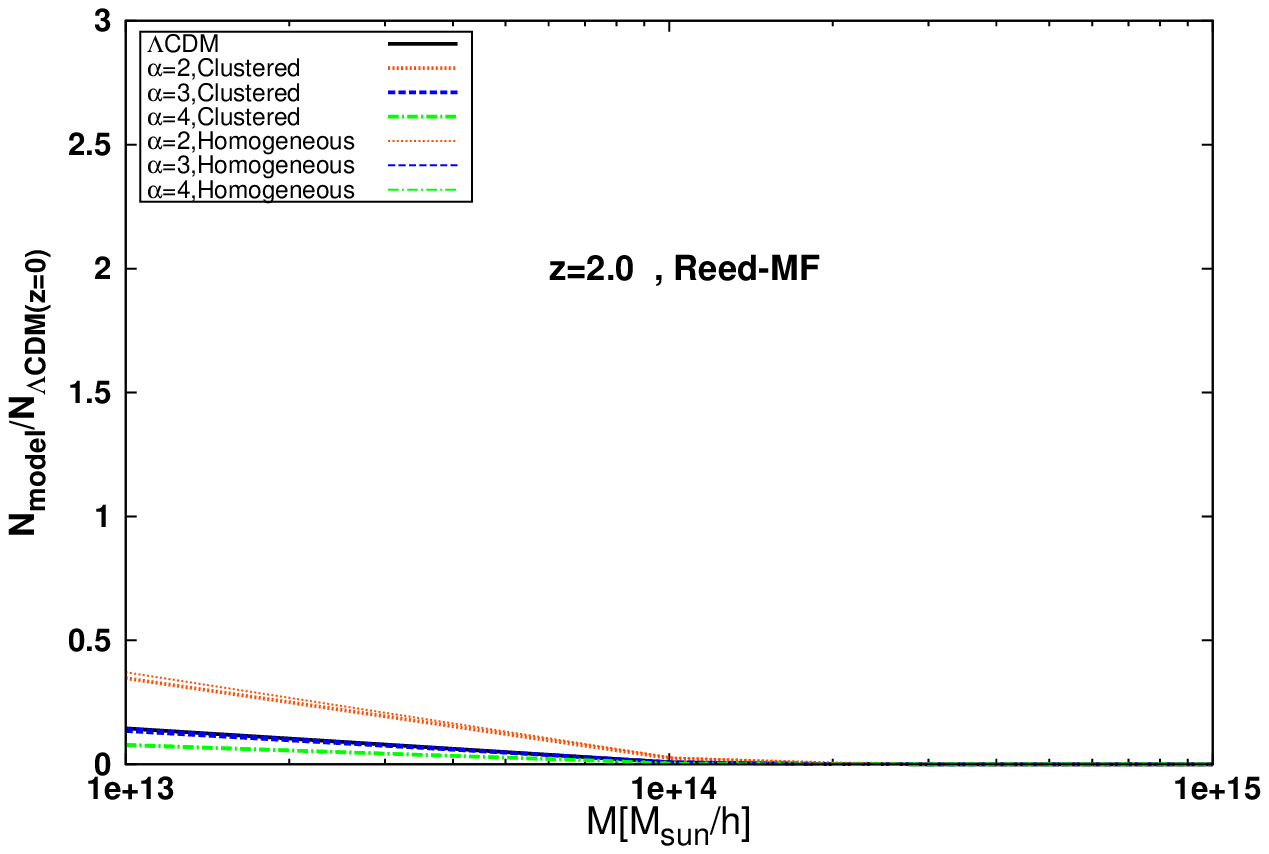}
	\caption{Ratio of the number density of cluster-size halos above a given mass $M$ for different NADE models  to the concordance $\Lambda$CDM cosmology at $z=0$ (first row panels), $z=0.5$ (second row panels), $z=1.0$ (third row panels) and $z=2.0$ (fourth row panels).  Line styles and colors are the same as in Fig.\ref{fig:d}.}
	\label{fig:nst}
\end{figure*}

 We also examine how the predicted number of halos are sensitive to the chosen mass function. To do this, we repeat our analysis using the Reed  mass function provided by \cite{Reed:2006rw}. In the Reed mass function, the authors fit their simulation data by steepening the high
mass slope of the Sheth-Tormen mass function by adding  new parameters $c$ and $G_1$ described as follows \citep{Reed:2006rw}:

\begin{equation}\label{reed}
f(\nu)=0.2709\sqrt{\dfrac{2}{\pi}}(1+1.1096\nu^{0.6}+0.2 G_1)exp(-\dfrac{0.707 c  \nu^2}{2})\;,
\end{equation}
where  $c=1.08$ and 
\begin{equation}\label{g1}
G_1=\exp (-\dfrac{(\ln \sigma^{-1}-0.4)^2}{2(0.6)^2})\;.
\end{equation}
 In Fig. \ref{fig:nst} we present the numerical results of our analysis by computing the number density of cluster-size halos at different redshifts: $z=0.0, 0.5, 1.0$ and $2.0$ for
three different values of NADE model parameter $\alpha$ considered in this work. To have a better comparison between all models, we normalize the results of NADE by that of the $\Lambda$CDM cosmology at $z=0$. The main results is sorted out as follows

\begin{figure*} 
	\centering
	\includegraphics[width=0.68 \textwidth]{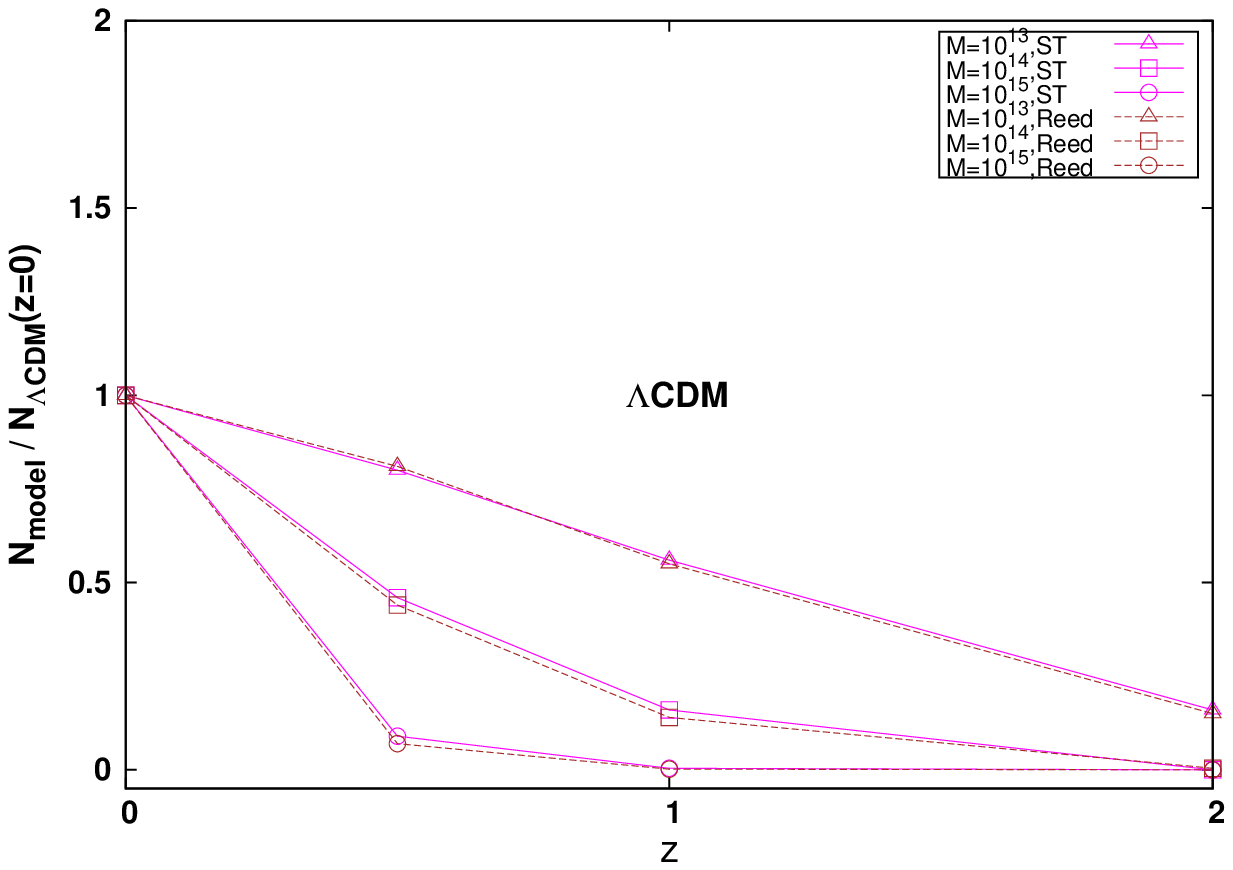}
	\includegraphics[width=0.32\textwidth]{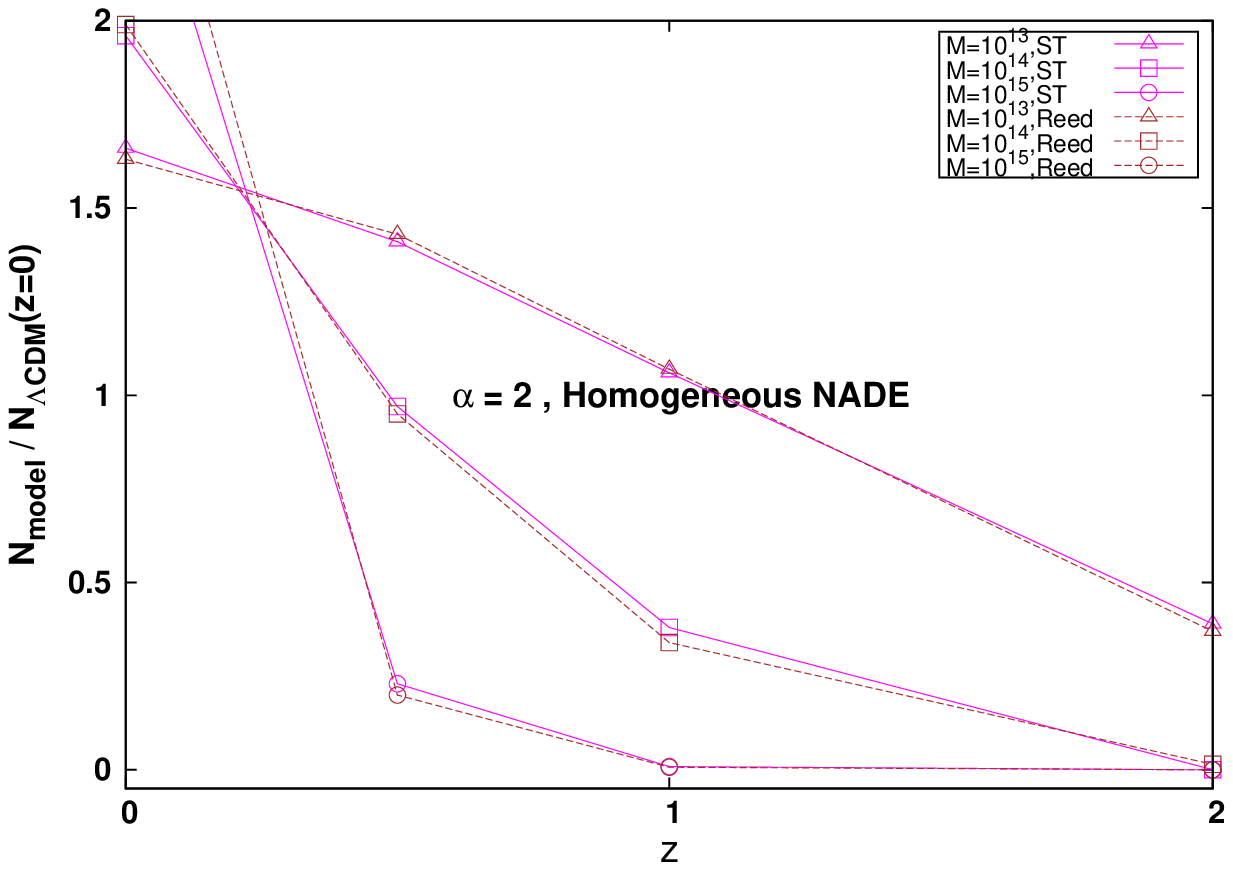}
    \includegraphics[width=0.32\textwidth]{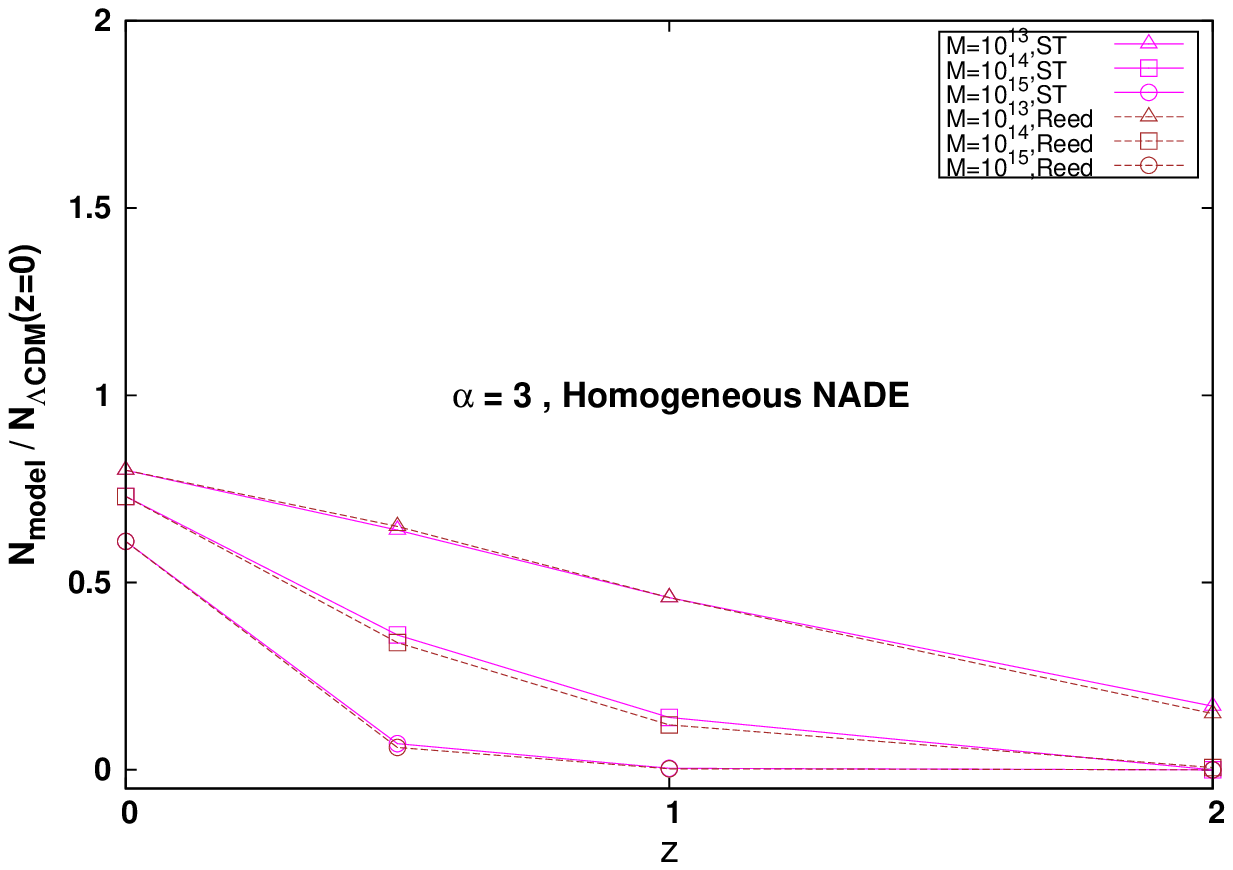}
    \includegraphics[width=0.32\textwidth]{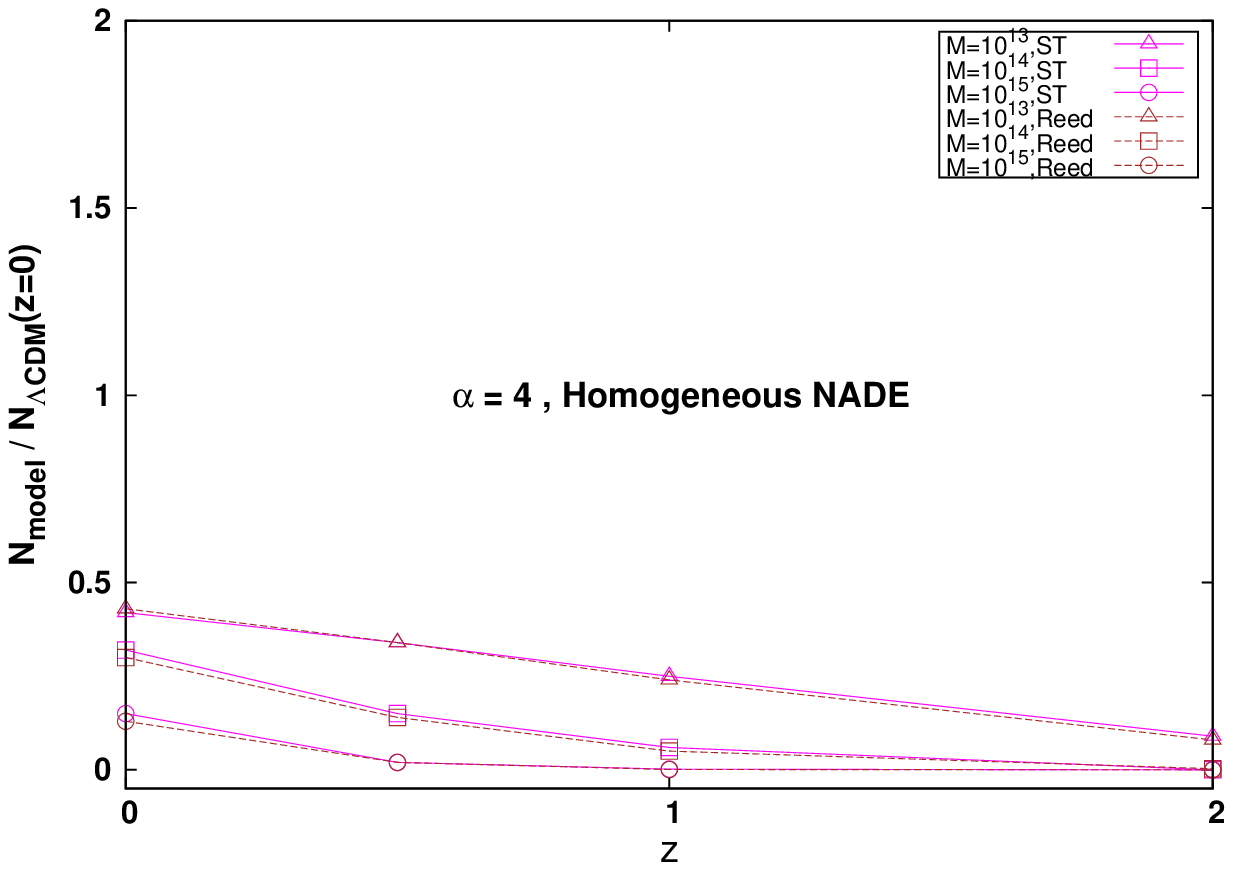}
	\includegraphics[width=0.32\textwidth]{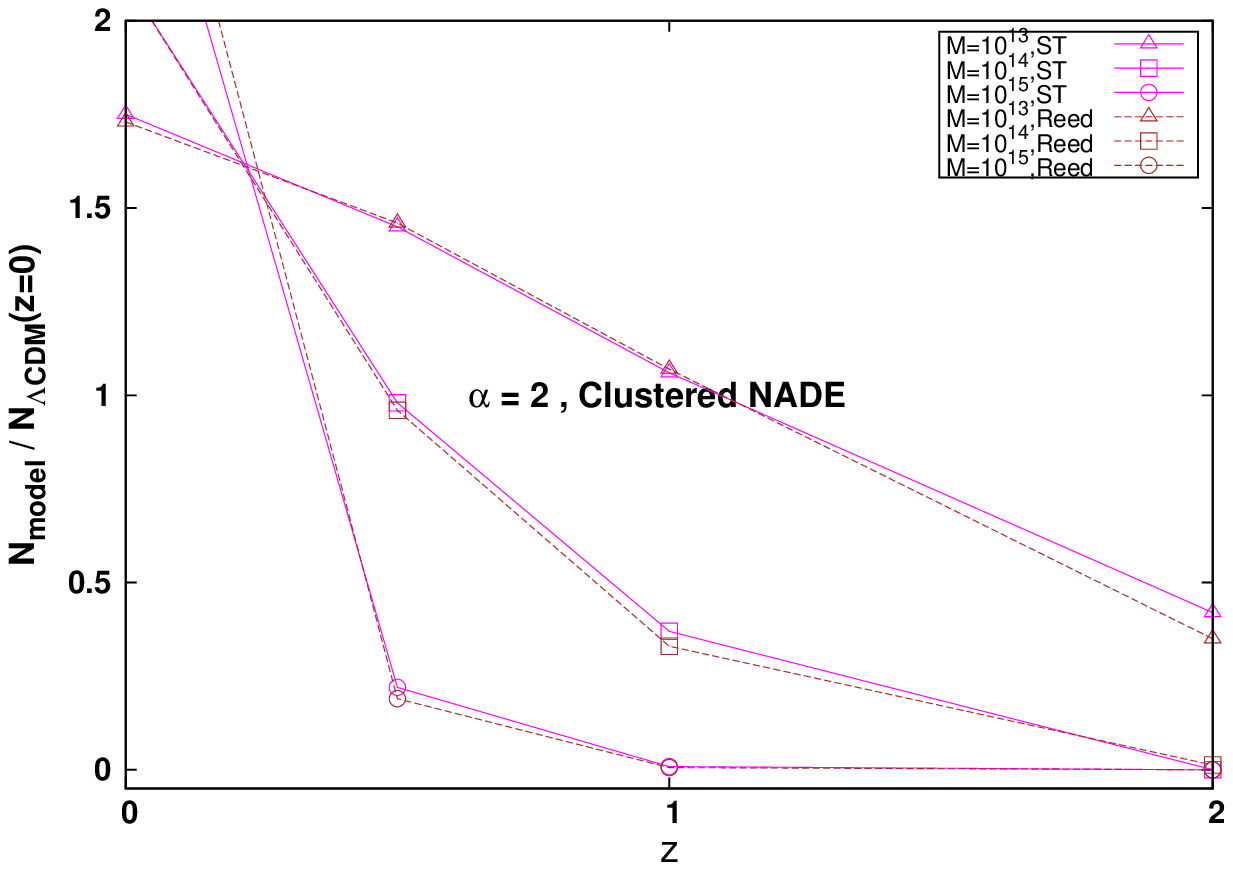}
    \includegraphics[width=0.32\textwidth]{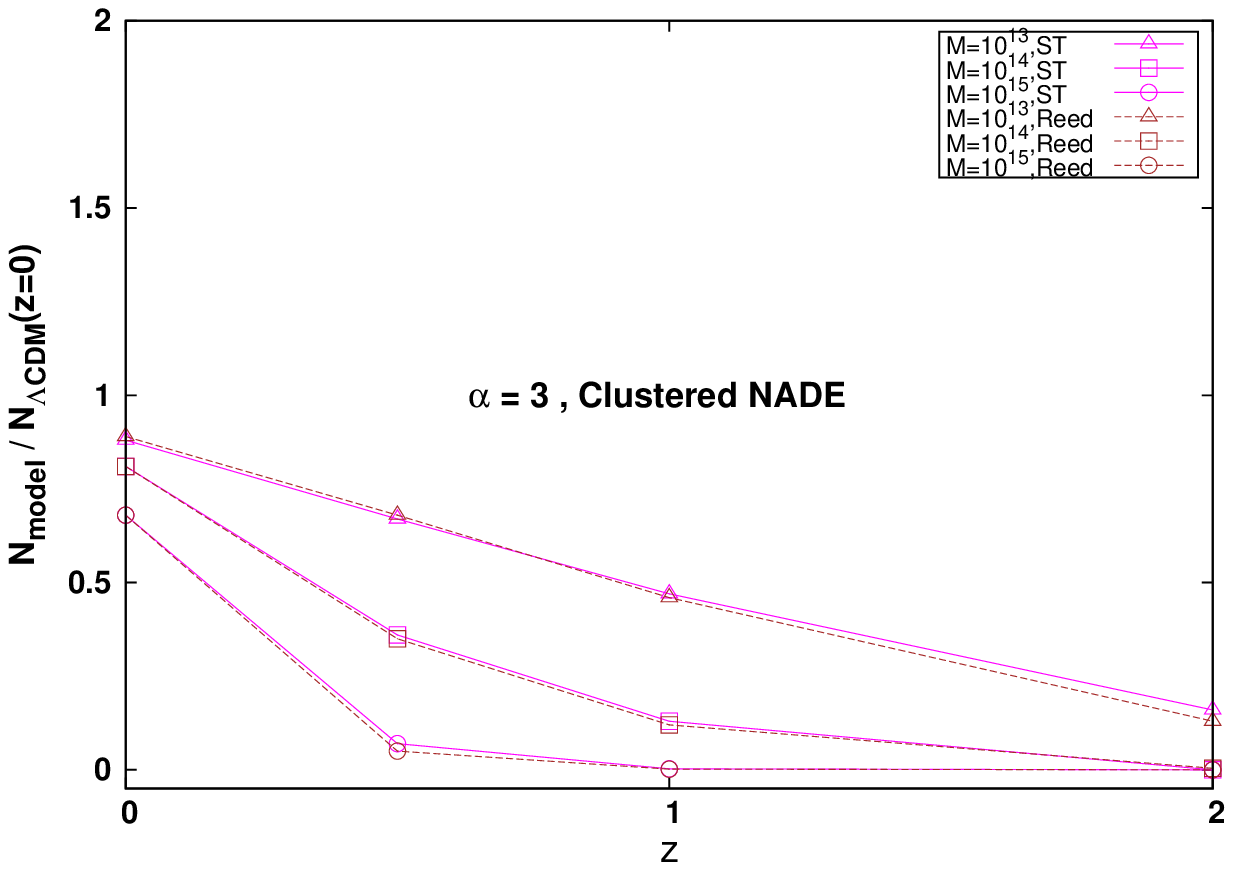}
    \includegraphics[width=0.32\textwidth]{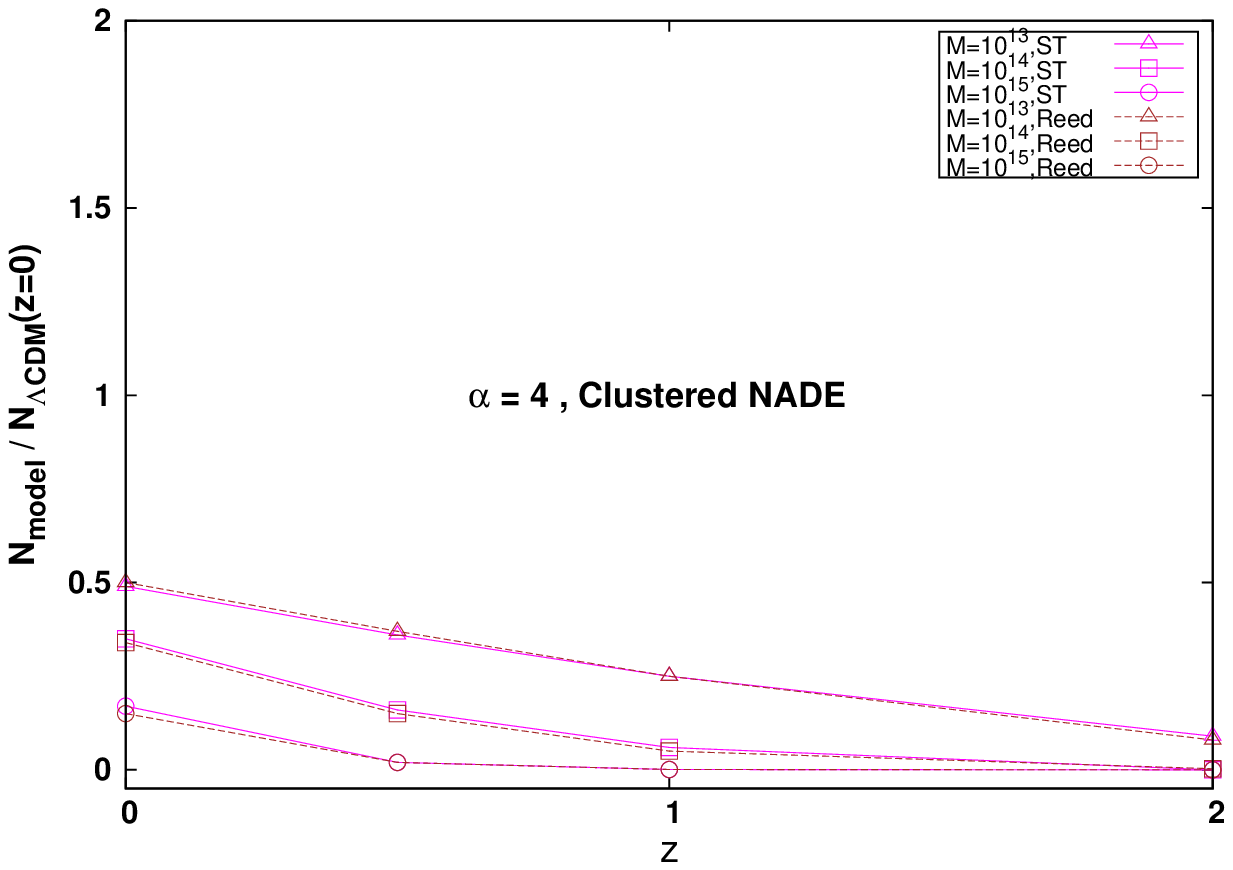}
	\caption{Redshift evolution of the number density of cluster-size halos normalized to that of the $\Lambda$CDM model calculated for different mass scales:$M>10^{13}M_{\rm sun}h^{-1}$, $M>10^{14}M_{\rm sun}h^{-1}$ and $M>10^{15}M_{\rm sun}h^{-1}$ in NADE and $\Lambda$CDM models. Line styles and colors  for ST and Reed mass functions are shown in the legends. }
	\label{fig:nz}
\end{figure*}

At $z=0$, for both Sheth-Tormen  and Reed mass functions, one can observe that in the cases $\alpha=3$ and $\alpha=4$ ($\alpha=2$) the NADE cosmology predicts less (more) abundance of
halos in comparison with the $\Lambda$CDM model at both the low and high mass tails. The similar results are achieved at $z=0.5$. The precise numerical results of our analysis for three different mass scales are presented in Table \ref{tab:nZ0} (see also Fig. \ref{fig:nz}). We observe that at $z=0$, the difference between NADE and $\Lambda$CDM models is considerable  at both low and high mass tails of the mass function. However, this difference is more pronounced for high mass ranges. Also the difference between the results of the two mass functions used appears in the high mass tail of clusters for the NADE model with $\alpha=2$.  In particular, in the case of $\alpha=2$, the number density of clusters with mass above $M=10^{15}M_{\rm sun}h^{-1}$ counted using the Reed mass function at $z=0$ is roughly $7\%$ higher than that of the ST mass function.

Moreover, at $z=0$ the clustered NADE models result somewhat more abundance of halos compared to homogeneous cases, while the difference is negligible at higher redshifts. Quantitatively speaking, the number density of halos with mass larger than $10^{13}M_{\rm sun}h^{-1}$ calculated at $z=0$ for the clustered NADE model with $\alpha=2$ is almost $5\%$ higher than the homogeneous case with the same $\alpha$.

 For all models, we see that by increasing the redshift $z$, the number density of clusters decreases.  Using the results presented in Table \ref{tab:nZ0}, we visualize the predicted number densities for three different mass scales: $M>10^{13}M_{\rm sun}h^{-1}$, $M>10^{14}M_{\rm sun}h^{-1}$ and $M>10^{13}M_{\rm sun}h^{-1}$  in Fig. \ref{fig:nz}. For example in the case of the standard $\Lambda$CDM model, the predicted number density of halos above $10^{13}M_{\rm sun}h^{-1}$ calculated using the ST mass function at $z=2$ is roughly $84\%$ lower than $z=0$. Notice that for all models,  the number density of massive halos with mass higher than  $10^{15}M_{\rm sun}h^{-1}$ at $z=2$ is roughly negligible compared to $z=0$. The above result tells us that the dark matter halos with smaller masses form sooner than larger ones. Moreover, we can conclude that the suppression effects of DE on the virializaion of halos are more pronounced in halos with higher masses. The same results are also found for Reed mass function.
 
\begin{table*}
 \centering
 \caption{Ratio of the number of cluster-size halos above given mass $M$ for different NADE models at different redshifts to the concordance $\Lambda$CDM cosmology at $z=0$,
 }
\begin{tabular}{c c c c c c c c c c}
\hline \hline
$z$ & $M[M_{\rm sun}/h]$  & MF &$\Lambda$CDM & \multicolumn{3}{|c}{Homogeneous NADE} &  \multicolumn{3}{|c}{Clustered NADE}\\
& &  & & $\alpha=2$ & $\alpha=3$ & $\alpha=4$& $\alpha=2$ & $\alpha=3$ & $\alpha=4$\\
 \hline
& & ST & 1.0 & 1.66 & 0.80 & 0.42 & 1.75 & 0.88 & 0.49 \\
 
$z=0$&$10^{13}$ &  & & & & & & & \\
 
  & & Reed & 1.0 & 1.63 & 0.80 & 0.43 & 1.73 & 0.89 & 0.50 \\
 \\\hline
 & & ST & 1.0 & 1.96 &0.73 & 0.32 & 2.11 & 0.81 & 0.35 \\
 
$z=0$ &$10^{14}$ &  & & & & & & & \\
 
 & & Reed & 1.0 & 1.99 &0.73 & 0.30 & 2.11 & 0.81 & 0.34 \\
\\\hline
&  & ST & 1.0 & 2.59 & 0.61 & 0.15 & 2.74 & 0.68 & 0.17 \\
 
$z=0$& $10^{15}$ &  & & & & & & & \\
 
 & & Reed & 1.0 & 2.79 & 0.61 & 0.13 & 2.94 & 0.68 & 0.15 \\
 \hline \hline
 & & ST & 0.80 & 1.41 & 0.64 & 0.34 & 1.45 & 0.67 & 0.36 \\
 
$z=0.5$&$10^{13}$ &  & & & & & & & \\
 
  & & Reed & 0.81 & 1.43 & 0.65 & 0.34 & 1.46 & 0.68 & 0.37 \\
\\\hline
 & & ST & 0.46 & 0.97 & 0.36 & 0.15 & 0.98 & 0.36 & 0.16 \\
 
$z=0.5$ &$10^{14}$ &  & & & & & & & \\
 
 & & Reed & 0.44 & 0.95 & 0.34 & 0.14 &0.96 & 0.35 & 0.15 \\
\\\hline
&  & ST & 0.09 & 0.23 & 0.07 &0.02 & 0.22 & 0.07 & 0.02 \\
 
$z=0.5$& $10^{15}$ &  & & & & & & & \\
 
 & & Reed & 0.07 & 0.20 & 0.06& 0.02 & 0.19 & 0.05 & 0.02 \\
  \hline \hline
 & & ST & 0.56 & 1.06 &0.46&0.25 & 1.06 & 0.47 & 0.25 \\
 
$z=1.0$&$10^{13}$ &  & & & & & & & \\
 
  & & Reed & 0.55 & 1.07 & 0.46 & 0.24 & 1.07 & 0.46 & 0.25 \\
\\\hline
 & & ST & 0.16 &0.38& 0.14 & 0.06& 0.37 & 0.13 & 0.06 \\
 
$z=1.0$ &$10^{14}$ &  & & & & & & & \\
 
 & & Reed & 0.14 & 0.34 & 0.12 & 0.05 & 0.33 & 0.12 & 0.05\\
\\\hline
&  & ST & 0.004 & 0.009 & 0.004 & 0.002 & 0.009 & 0.003 & 0.001 \\
 
$z=1.0$& $10^{15}$ &  & & & & & & & \\
 
 & & Reed & 0.002 & 0.007 & 0.003 &0.001 & 0.006 & 0.002 & 0.001 \\
  \hline\hline
 & & ST & 0.16 & 0.39 & 0.17 & 0.09 & 0.42 & 0.16 & 0.09 \\
 
$z=2.0$&$10^{13}$ &  & & & & & & & \\
 
  & & Reed & 0.15 & 0.37 & 0.15 & 0.08 & 0.35 & 0.13 & 0.08 \\
\\\hline
 & & ST & $5 \times 10^{-5}$ &$3 \times 10^{-4}$ &$ 2\times 10^{-4}$ &$1\times 10^{-4} $& $4\times 10^{-4}$ & $1\times 10^{-4}$ &$ 8\times 10^{-5}$ \\
 
$z=2.0$ &$10^{14}$ &  & & & & & & & \\
 
 & & Reed & 0.004 & 0.015 & 0.006 & 0.003 & 0.013 & 0.004 & 0.003 \\
\\\hline
&  & ST & $1 \times 10^{-7} $& $1 \times 10^{-6} $&$ 2 \times 10^{-6}$ & $1 \times 10^{-6}$ &$ 2 \times 10^{-6}$ &$ 8 \times 10^{-7}$&$ 6 \times 10^{-7}$ \\
 
$z=2.0$& $10^{15}$ &  & & & & & & & \\
 
 & & Reed & $4 \times 10^{-8}$ &$ 3 \times 10^{-7}$ &$3 \times 10^{-7}$ & $3 \times 10^{-7}$ & $2 \times 10^{-7}$ &$ 2 \times 10^{-7} $&$ 2 \times 10^{-7}$ \\
 \hline \hline
\end{tabular}\label{tab:nZ0}
\end{table*}

\section{Conclusion}\label{conclude}

In this work we studied the SCM and predicted the number of dark matter halos in the framework of NADE cosmologies. We first studied the evolution of Hubble expansion in this model. We saw that the EoS parameter of NADE  remains in the quintessence regime and cannot cross the phantom line.

Then we studied the impact of DE in the NADE model on the collapse of dark matter halos in the framework of the SCM. In particular, the effects of DE on the linear growth factor of perturbations, ISW, the linear and virial overdensities and  the abundance of virialized halos were investigated. 

While DE accelerates the expansion rate of Hubble flow, it has two different rules on the formation of cosmic structures. In the framework of homogeneous NADE, DE suppresses the growth of dark matter perturbations. On the other hand, in the case of clustered NADE, DE perturbations can enhance the growth of matter fluctuations. Depending on the model parameter $\alpha$, the growth factor of perturbations $D_+$ can be larger or smaller than standard $\Lambda$CDM cosmology. Notice that NADE for all the values of $\alpha$ results the higher growth factor compared to an EdS universe.

Measuring the ISW effect as a useful observational tool, we showed that depending on $\alpha$ and redshift $z$  this effect in NADE cosmologies can be smaller or larger than that in concordance $\Lambda$CDM cosmology. We also showed that the ISW effect in clustered NADE models is somewhat larger than the homogeneous cases.

The two main parameters of SCM, $\delta_{\rm c}$ and $\Delta_{\rm vir}$, have been computed. Similar to what happened for growth factor $D_+$ and the ISW effect, we saw that the evolution of these quantities strongly depends on the model parameter of NADE such that $\delta_{\rm c}$ and $\Delta_{\rm vir}$ become smaller for larger values on $\alpha$ at low redshifts. In particular, we conclude that the low dense virialized halos can be formed for higher values of $\alpha$.

We computed the predicted number of virialized dark matter halos using the two relevant Sheth-Tormen and Reed mass functions in the context of clustered and homogeneous NADE models respectively. Notice that in the case of clustered NADE model, we used the corrected  mass function formula by adding the contribution of the DE mass on the total mass of clusters. It has been shown that the abundance of halos at different redshifts depends on the model parameter $\alpha$ of NADE cosmologies.
We showed our results for four different redshifts $z=0, 0.5, 1.0$ and $2.0$ and saw that for all mentioned redshifts both mass functions predict a grater abundance of halos in NADE cosmology for $\alpha=2$ compared to the $\Lambda$CDM universe. For higher values $\alpha=3$ and $\alpha=4$, we observe fewer abundant halos in NADE compared to the $\Lambda$CDM until $z\lesssim1$. Along the redshift, the number density of halos computed in our analysis is decreasing. These decrements are more pronounced for massive halos compared to low-mass objects. This result is compatible with the fact in standard gravity that the low mass dark matter halos form sooner than the larger ones.  Also the suppression effects of DE in NADE cosmology on the virializaion of cluster-size halos are more significant at higher masses. It has been shown that all qualitatively results obtained whit the Sheth-Tormen mass function are also valid in the Reed mass function. We also concluded that the number of dark matter halos computed at low redshifts in clustered NADE cosmology is higher than that of homogeneous cases. Notice that at high redshift $z=2$ where the abundance of halos falls down, the differences between clustered and homogeneous models become negligible.


 \bibliographystyle{apsrev4-1}
  \bibliography{ref}

\end{document}